\documentclass[useAMS,usenatbib,pdflscape]{mn2e}
\usepackage{graphicx}
\usepackage{epstopdf}
\usepackage{amsmath}
\usepackage{amssymb}
\usepackage{rotating}
\usepackage{pdflscape}

\title[]{Effect of nuclear stars gravity on quasar radiation feedback on the parsec-scale}
\author[Yang \& Bu]{Xiao-Hong Yang$^1$\thanks{E-mail: yangxh@cqu.edu.cn (XH)} and De-Fu Bu$^2$\thanks{E-mail: dfbu@shao.ac.cn (DB)}\\
$^1$ Department of Physics, Chongqing University, Chongqing 400044, China\\
$^{2}$Key Laboratory for Research in Galaxies and Cosmology, Shanghai Astronomical Observatory, Chinese Academy of Sciences,\\ 80 Nandan Road, Shanghai 200030, China}

\begin{document}

\pagerange{\pageref{firstpage}--\pageref{lastpage}} \pubyear{20**}

\maketitle

\label{firstpage}
\def\LSUN{\rm L_{\odot}}
\def\MSUN{\rm M_{\odot}}
\def\RSUN{\rm R_{\odot}}
\def\MSUNYR{\rm M_{\odot}\,yr^{-1}}
\def\MSUNS{\rm M_{\odot}\,s^{-1}}
\def\MDOT{\dot{M}}

\begin{abstract}
It is often suggested that a super massive black hole is embedded in a nuclear bulge of size of a few $10^2$ $\text{ parsec}$ . The nuclear stars gravity is not negligible near $\sim 10 \text{ parsec}$. In order to study the effect of nuclear stars gravity on quasar radiation feedback on the parsec scale, we have simulated the parsec scale flows irradiated by a quasar by taking into account the gravitational potential of both the black hole and the nuclear star cluster. We find that the effect of nuclear stars gravity on the parsec-scale flows is related to the fraction of X-ray photons in quasar radiation. For the models in which the fraction of X-ray photons is not small (e.g. the X-ray photons contribute to $20\%$ of the quasar radiation), the nuclear stars gravity is very helpful to collimate the outflows driven by UV photons, significantly weakens the outflow power at the outer boundary and significantly enhances the net accretion rate onto the black hole. For the models in which X-ray photons are significantly decreased (e.g. the X-ray photons contribute to $5\%$ of the quasar radiation), the nuclear stars gravity can just slightly change properties of outflow and slightly enhance the net accretion rate onto the black hole.

\end{abstract}


\section{Introduction}
A super massive black hole (SMBH) is commonly suggested to be located at the centre of galaxies. The SMBH can accrete matter from its vicinity and power active galactic nuclei (AGNs) to emit electromagnetic energy from radio to X-rays and beyond. Therefore, AGNs, quasars in particular, are very powerful radiation sources. The AGNs strong radiation influences the ionization structure and dynamics of matters not only in the vicinity of AGNs but also on larger scales. The AGNs feedback effects are thought to play an important role in the processes of galaxy formation and evolution. In the past decade, it is observed that the SMBH mass ($M_{\rm BH}$) is tightly correlated with host galaxy properties (e.g. Ferrarese \& Merritt 2000; Gebhardt et al. 2000; Graham et al. 2001; Greene \& Ho 2006; Kormendy \& Bender 2009; Graham \& Scott 2015).

AGNs feedback effects on different scales have been numerically studied in previous works (e.g. Ciotti \& Ostriker 1997; 2001; 2007; Brighenti \& Mathews 2003; Hopkins et al. 2005; Springel et al. 2005, Sazonov et al. 2005; Proga 2007; Proga et al. 2008; Kurosawa \& Proga 2009; Ciotti, Ostriker \& Proga 2009, 2010; Novak, Ostriker \& Ciotti 2012; Liu et al. 2013; Gan et al. 2014), but this issue is still open to date. The region from the outer boundary of black hole accretion disc to a few tens parsec (pc) is an important domain, through which not only matters are accreted into the BH accretion disc as fuels but also the radiation and mechanical feedback of AGNs are passed to the galaxy scale. A series of works are devoted to studying the radiation feedback of quasars and low-luminosity AGNs (LLAGNs) on the sub-pc and pc scales (Proga 2007; Proga et al. 2008; Kurosawa \& Proga 2009; Yuan \& Li 2011; Liu et al. 2013; Yang \& Bu 2018; Bu \& Yang 2018). LLAGNs and quasars emit different spectra, which cause different radiation feedback effects. For LLAGNs, a great number of high-energy photons (mainly in X-ray band) from the innermost region of accretion flows can heat the gas at the Bondi radii. If the radiation is strong enough, the gas temperature at the Bondi radius will be heated to be above the Virial temperature, so that outflows will be produced, the accretion process will be stopped, and then the intermittent activity will take place (Bu \& Yang 2018). Such radiation feedback mechanism was proposed by Yuan \& Li (2011) as the origin of the intermittent activity of some compact young radio sources with $L\ga 2\%L_{\rm Edd}$. For a quasar, most of the emitted photons are in the optical-UV bands, but some fraction is in the X-ray band. X-ray photons heat and ionize the gas around AGNs, whereas UV photons drive away the gas by electron scattering and spectral lines froce (i.e. line force). Barai et al. (2011; 2012) and Moscibrodzka \& Proga (2013) have considered only X-ray photons in their simulations and suggested that the feedback due to X-ray heating is important to the thermal and dynamical properties of gas accreting onto the SMBH. The feedback effects of quasar radiation (including X-ray and UV photons) have been numerically simulated by Proga and his collaborators (Proga 2007; Proga et al. 2008; Kurosawa \& Proga 2009; Liu et al. 2013), whose study clearly demonstrates that quasar radiation can drive nonspherical, multitemperature, and very dynamic flows. This mechanism can be used to explain some highly blueshifted broad absorption line features seen in the UV and optical spectra of AGNs (e.g. Murray et al. 1995).

In reality, physical processes on sub-pc and pc scales may be complex. It is suggested that dusts can survive outside of the sublimation radius $R_{\rm sub}\simeq0.54(\frac{L}{10^{46} \text{ erg s}^{-1}})^{1/2}(\frac{T}{1500 \text{ K}})^{-2}$ pc, where $L$ is the luminosity of black hole accretion disc. The radiation pressure on dusts can play a key role in the dynamics of matters near $\sim$1 pc (Dorodnitsyn et al. 2016). A mixture of gas and dust should be taken into account on pc scale. The winds produced from accretion disc have been found in observations and numerical simulations(Murray et al. 1995; Proga, Stone \& Kallman 2000; Tombesi et
al. 2011; Gofford et al. 2013; Nardini et al. 2015). The winds can also influence the dynamics of gas on pc scale.

In addition, observations show that a SMBH of $\sim10^8-10^9 M_{\bigodot}$ is embedded in a nuclear bulge of size of a few $10^2$ pc (e.g. Philipp et al. 1999). The BH mass appears always to be a fairly constant fraction of the stellar bulge mass $M_{\rm b}$, i.e. $M_{\rm BH}\sim10^{-3}M_{\rm b}$ (H\"{a}ring \& Rix 2004), and observations indicate a tight relation of the BH mass and the velocity dispersion ($\sigma_{\ast}$) of stars in bulge (Ferrarese \& Merritt 2000; Gebhardt et al. 2000; Greene \& Ho 2006). There is the radius of ``sphere of influence'' of the BH gravity, which reads $r_{\rm inf}=GM_{\rm BH}/\sigma_{\ast}^2\simeq 8M_{8}/\sigma_{200}^2 \text{ pc}$, where $M_{8}=M_{\rm BH}/10^8M_{\bigodot}$ ($M_{\bigodot}$ is solar mass) and $\sigma_{200}=\sigma_{\ast}/200 \text{ km s}^{-1}$ (e.g. King \& Pounds 2015). Within $r_{\rm inf}$, the BH potential significantly affects the motion of the stars or of the interstellar medium. Outside of $r_{\rm inf}$, the impact of BH gravity on the global stellar and interstellar medium dynamics is significantly reduced. The $M_{\rm BH}-\sigma_{\ast}$ correlation demonstrates that the velocity dispersion of stars is a constant, which seems to be the case for many AGNs ( e.g. Kormendy \& Ho 2013). So the gravitational potential of the nuclear star clusters is assumed to be $\propto\sigma_{\ast}^2\text{ln}(r)$ within the nuclear bulge (Bu et al. 2016). In this case, in addition to the potential of the BH, the gravitational potential of the nuclear star cluster will become important on the pc-scale and should be included. At larger scale (e.g. galaxy scale), the gravitational potential of host galaxies should be included. For example, the Bondi accretion at galaxy scale have been explored (Barai et al. 2011; Korol, Ciotti \& Pellegrini 2016; Ciotti \& Pellegrini 2017). In this paper, we will focus on effect of nuclear stars gravity on quasar radiation feedback on pc scale.

We organize our paper as follows: Section 2 describes our models and numerical method; Section 3 presents the simulation results; Section 4 is devoted to a summary.

\section{METHOD}
\subsection{Basic equations}
The method of the calculation is similar to that of Proga (2007), Proga et al. (2008), and Kurosawa \& Proga (2009). We consider that the gases at $\sim 1 \rm pc$ scale are irradiated by a quasar. In a spherical coordinate $(r, \theta, \phi)$, we use the ZEUS-MP code (Hayes et al. 2006) to solve the following hydrodynamical equations:
\begin{equation}
\frac{d\rho}{dt}+\rho\nabla\cdot \mathbf{v}=0,\label{cont}
\end{equation}
\begin{equation}
\rho\frac{d\mathbf{v}}{dt}=-\nabla P-\rho\nabla
\psi+\rho {\bf F}^{\rm rad}, \label{monentum}
\end{equation}
\begin{equation}
\rho\frac{d(e/\rho)}{dt}=-P\nabla\cdot\mathbf{v}+\rho \cal{L}.
\end{equation}
Here, $d/dt(\equiv \partial / \partial t+ \mathbf{v} \cdot \nabla)$ denotes the Lagrangian time derivative; $\rho$ , $P$, $\mathbf{v}$, $\psi$ and $e$ are density, pressure, velocity, gravitational potential and internal energy, respectively; ${\bf F}^{\rm rad}$ is the total radiation force per unit mass and $\cal{L}$ is the net cooling rate. We adopt an adiabatic equation of state of ideal gas $P=(\gamma -1)e$ and set the adiabatic index $\gamma =5/3$.

\begin{figure}

\scalebox{0.5}[0.5]{\rotatebox{0}{\includegraphics[bb=50 40 531
436]{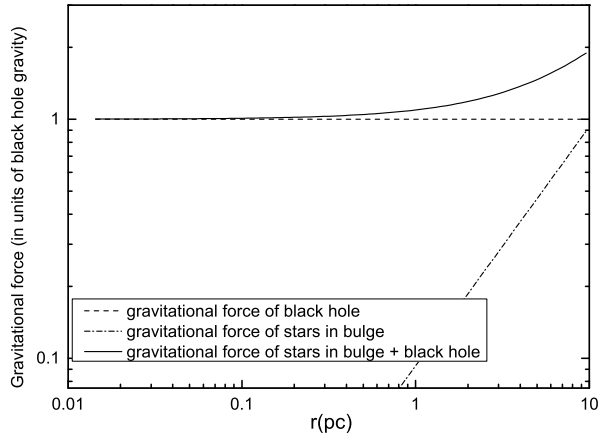}}}
\ \centering \caption{Distribution of gravitational force.}\label{fig 1}
\end{figure}

The radiation force ${\bf F}^{\rm rad}$ includes the Compton scattering force and the line force produced by the UV photons coming from a quasar. It is simply assumed that the electromagnetic emissions of quasar are produced by the most inner part of the accretion disc plus an extended corona. The most inner part of the accretion disc emits only the UV photons whose luminosity $L_{\rm UV}=f_{\rm UV} L$, whereas the corona emits only X-rays whose luminosity $L_{\rm X}=f_{\rm X} L=(1-f_{\rm UV})L$. Here, $L$ is the total luminosity of the quasar, $f_{\rm UV}$ is the ratio of UV luminosity to the total luminosity and $f_{\rm X}$ is the ratio of X-rays luminosity to the total luminosity, respectively. We ignore the emissions in the remaining bands, mainly optical and infrared. In our calculation, the inner boundary of the computational domain is much larger than the radius of UV and X-ray emission source. Therefore, the emission source is simply approximated as a point source, so that ${\bf F}^{\rm rad}$ has only the radial component. Following Kurosawa \& Proga (2009), the radial component of ${\bf F}^{\rm rad}$ reads
 \begin{equation}
{\bf F}^{\rm rad}_r=\frac{\sigma_{\rm e} L}{4\pi r^2 c}[f_{\rm X}+2 cos(\theta)f_{\rm UV}(1+\mathcal{M})],
\end{equation}
where $\sigma_{\rm e}$ is the election-scattering cross section and $\mathcal{M}$ is the force multiplier describing the radiative acceleration resulting from spectral lines (i.e. line force) (Stevens \& Kallman 1990). The force multiplier $\mathcal{M}$ is a function of the ionization parameter ($\xi$) and the local optical depth parameter (cf. Rybichi \& Hummer 1978). For More details of $\mathcal{M}$, readers are referred to Proga (2007), where the force multiplier also depends formally on the thermal speed ($v_{\rm th}$) of gas. The thermal speed is set to 20 km s$^{-1}$, which corresponds the gas temperature of $2.5\times10^4$ K (Stevens \& Kallman 1990; Proga 2007).

\begin{table*}
\begin{center}

\caption[]{Summary of Models in section 3.1}

\begin{tabular}{ccccccccccc}
\hline\noalign{\smallskip} \hline\noalign{\smallskip}

Run &  $f_{\text{\rm UV}}$ & $f_{\rm X}$ &  $r'_{\rm c}$ & $\sigma_{\ast}$   & $\dot{M}_{\text{in}}$($6\times10^5 r_{s}$)  & $\dot{M}_{\text{out}}$($6\times10^5
r_{s}$) & $\dot{M}_{\text{net}}$($r_{\rm in}$) & $v_{\text{r}}$ & $P_{\text{k}}$ & $P_{\text{th}}$ \\
    &      & & ($r_{\ast}$) & ($100\text{ km s}^{-1}$) & ($\dot{M}_{\rm Edd}$) & ($\dot{M}_{\rm Edd}$) & ($\dot{M}_{\rm Edd}$) & ($\text{km s}^{-1}$) & ($L_{\rm Edd}$)
    & ($L_{\rm Edd}$) \\
(1) & (2)             & (3)                         &  (4)      &     (5) & {6}  & (7) & (8) & (9) & (10)   & (11)     \\

\hline\noalign{\smallskip}
B0   & 0.8  & 0.2     &0    & 0    &-0.45   &0.27  &-0.14 &3200 &1.1$\times10^{-4}$ &5.1$\times10^{-9}$ \\
B1   & 0.8  & 0.2     &0    & 2.0  &-0.53   &0.15  &-0.34 &1500 &2.0$\times10^{-5}$ &5.9$\times10^{-9}$ \\
\hline\noalign{\smallskip}
C0   & 0.95 &0.05     &0    & 0    &-0.53   &0.44  &-0.10 &1600 &5.8$\times10^{-5}$ &9.5$\times10^{-9}$ \\
C1   & 0.95 &0.05     &0    & 2.0  &-0.57   &0.49  &-0.10 &1300 &5.1$\times10^{-5}$ &7.6$\times10^{-9}$ \\

 \hline\noalign{\smallskip} \hline\noalign{\smallskip}
\end{tabular}
\end{center}

\begin{list}{}
\item\scriptsize{\textit{Note}. Columns (2), (3), (4): $f_{\rm UV}$, $f_{\rm X}$, the circularization radius $r'_{\rm c}$ and the star velocity dispersion,
    respectively; columns (6) and (7): the mass inflow rate ($\MDOT_{\rm in}$) and the mass outflow rate ($\MDOT_{\rm out}$) through the radius of $6\times10^5 r_{\rm
    s}$; columns (8): the net mass flux through the inner boundary ($\MDOT_{\rm in}$); columns (9): the maximum outflow velocity ($v_{r}$) at the outer boundary;
    columns (10) and (11): the kinetic energy ($P_{\rm k}$) and the thermal energy ($P_{\rm th}$) carried out by the outflow through the radius of $r=6\times10^5
    r_{\rm s}$.}
\end{list}
\label{tab1}
\end{table*}

For the X-ray irradiated gas, the net cooling rate $\cal{L}$ relates to the Compton and photoionization heating and the Compton, radiative recombination, bremsstrahlung, and line cooling. It depends on the density ($\rho$), the gas temperature ($T_{\rm g}$), the ionization parameter ($\xi$), and the characteristic temperature of the X-ray radiation ($T_{\rm X}$). Previous works have computed the analytical formulae of $\cal{L}$. Blondin (1994) considered an optically thin, cosmic abundance gas irradiated by a 10 kev bremsstrahlung spectrum and gave a series of approximate analytic formulae of $\cal{L}$. Their results were found to be in reasonable (25\%) agreement with numerical simulations. Dorodnitsyn et al. (2008) slightly modified Blondin's approximations using new atomic data. Similarly, Sazonov et al. (2005) have obtained another version of approximate formulae based on the XSTAR photoionization code (Kallman 2002). Here, we follow Proga, Stone \& Kallman (2000) and adopt Blondin's approximations. For convenience, we copy their results here (all quantities are expressed in the cgs system),

\begin{equation}
\rho {\cal{L}} = \textit{n}^2 (G_{\rm C}+G_{\rm X}-L_{\rm b,\textit{l}}),
\end{equation}
where, $\textit{n}$ is the number density of gas, $G_{\rm C}$ is the Compton heating/cooling rate,
\begin{equation}
G_{\rm C} = 8.9\times10^{-36} \xi (T_{\rm X}-4 T_{\rm g}),
\end{equation}
$G_{\rm X}$ is the rate of X-ray photoionization heating and recombination cooling,
\begin{equation}
G_{\rm X} = 1.5\times10^{-21} \xi^{1/4}T_{\rm g}^{-1/2} (1-T_{\rm g}/T_{\rm X}),
\end{equation}
$L_{\rm{b,l}}$ is the bremsstrahlung and line cooling,
\begin{eqnarray}
\nonumber L_{\rm b,\textit{l}} = 3.3\times10^{-27} T_{\rm g}^{1/2}+\delta[1.7\times10^{-18}\\
\times e^{(-1.3\times10^5/T_{\rm g})}\xi^{-1}T_{\rm g}^{-1/2}+10^{-24}].
\end{eqnarray}
Here, the ionization parameter $\xi=4\pi \textit{F}_{\rm X}/n$, where $\textit{F}_{\rm X}$ is the local X-ray flux, and the parameter $ \delta$ is introduced to this formula for controlling line cooling. $ \delta= 1$ represents the optically thin line cooling, whereas $\delta < 1$ indicates that the line cooling is reduced when the lines becomes optically thick. We set $ \delta= 1$.

The gravitational potential $\psi$ includes the contribution of the center black hole and the nuclear star clusters, so that $\psi$ can be expressed as
\begin{equation}
\psi= \psi_{BH}+\psi_{star}.
\end{equation}
The $\psi_{BH}$ denotes the gravitational potential of the center black hole, $\psi_{BH}=-GM_{BH}/r$, where $G$ is the gravitational constant. The $\psi_{star}$ denotes the gravitational potential of the nuclear star cluster. Bu et al. (2016) give that the potential of the star cluster is $\psi_{star}=\sigma_\ast^2 \ln (r)+C$, where $\sigma$ is the velocity dispersion of stars and $C$ is a constant. A typical physical value of $\sigma_{\ast}$ is $\sim (100-400) {\rm ~km~s^{-1}}$ (e.g., Kormendy \& Ho 2013). Greene \& Ho (2006) have shown that the stellar dispersion velocity $\sigma_{\ast}$ of bulges is about 150$-$250$\text{km s}^{-1}$ for a black
hole of $10^8 M_{\bigodot}$. Therefore, we consider the effect of stars gravity in bulges on the radiation feedback of quasars under the assumption of $\sigma_\ast=200 \text{ km s}^{-1}$. Figure 1 shows the gravitational force distribution. When $\sigma_\ast=200 \text{ km s}^{-1}$, the stars gravity in bulges significantly contributes to the sum of gravitational forces beyond $\sim 1 \text{ pc}$.

\begin{figure*}

\scalebox{0.29}[0.35]{\rotatebox{0}{\includegraphics[bb=70 340 480 700]{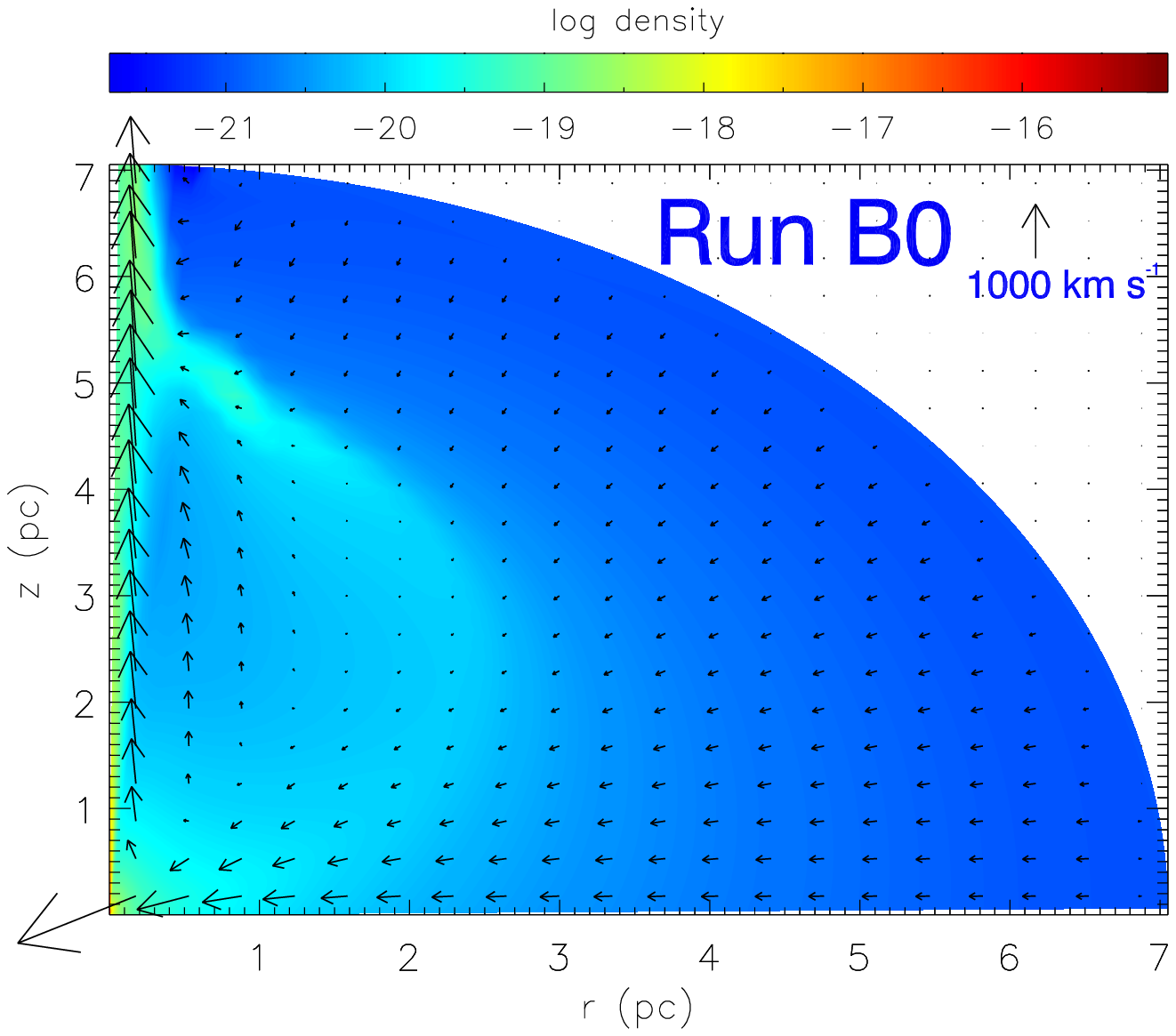}}}
\scalebox{0.29}[0.35]{\rotatebox{0}{\includegraphics[bb=70 340 480 700]{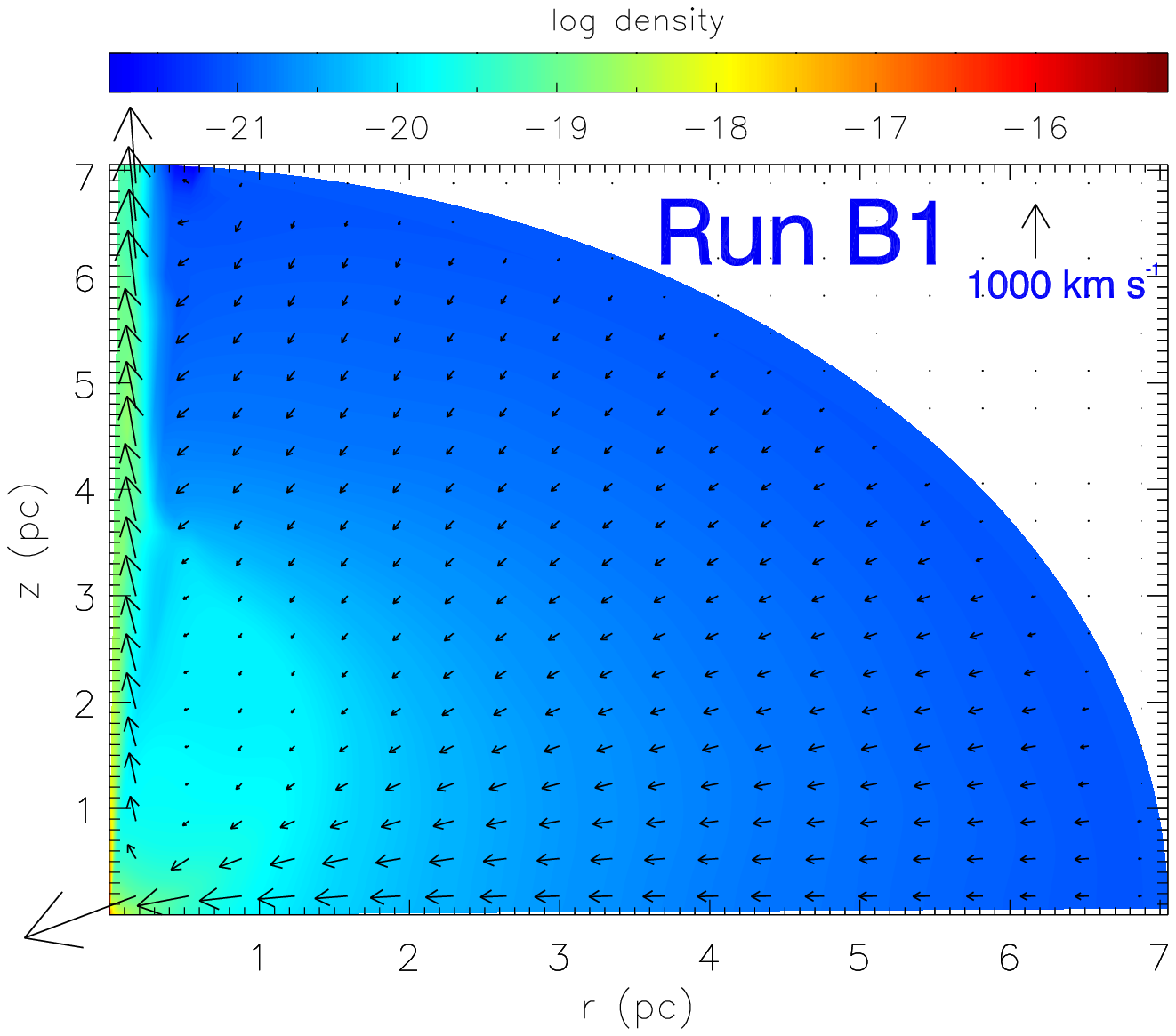}}}
\scalebox{0.29}[0.35]{\rotatebox{0}{\includegraphics[bb=70 340 480 700]{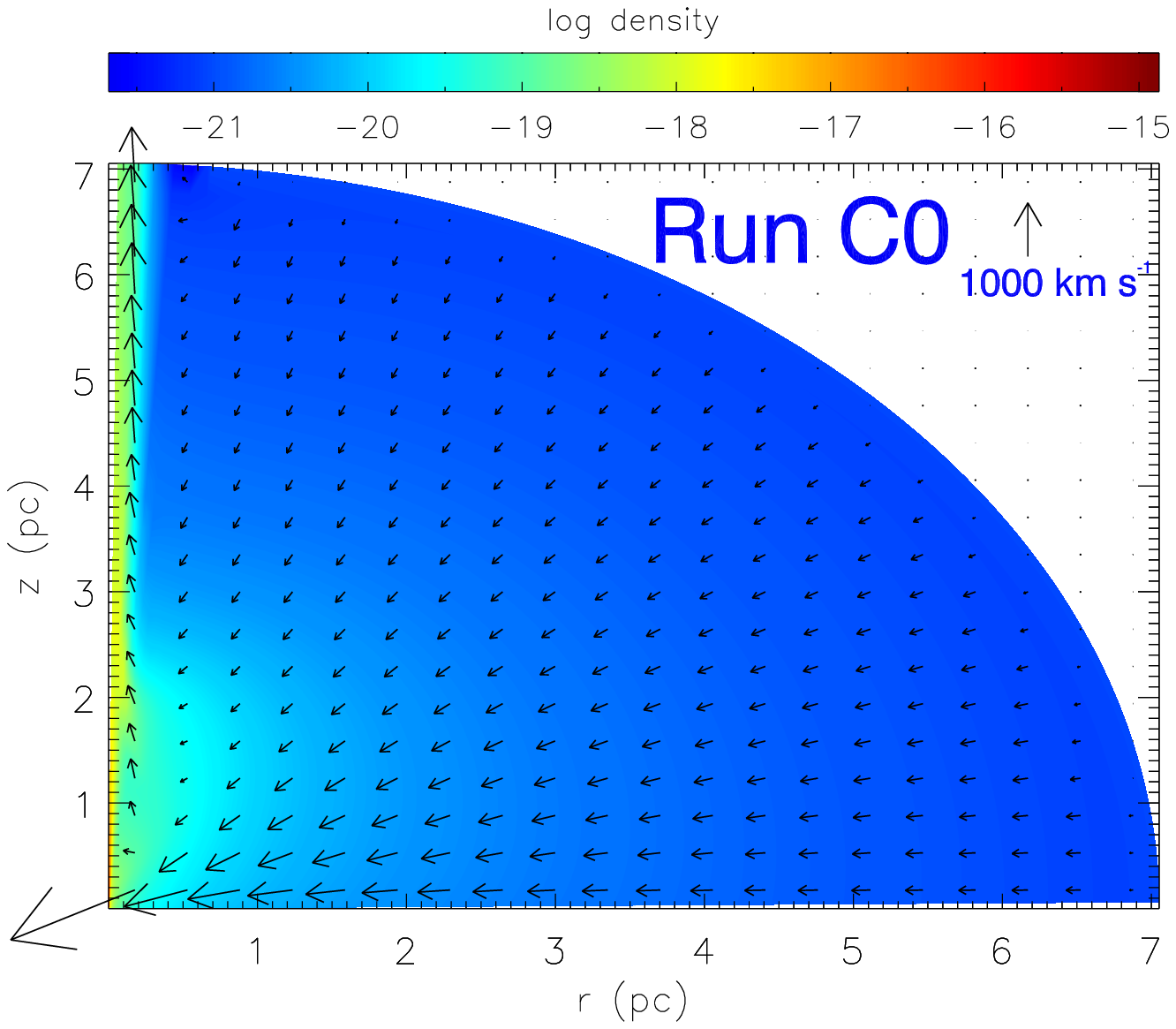}}}
\scalebox{0.29}[0.35]{\rotatebox{0}{\includegraphics[bb=70 340 480 700]{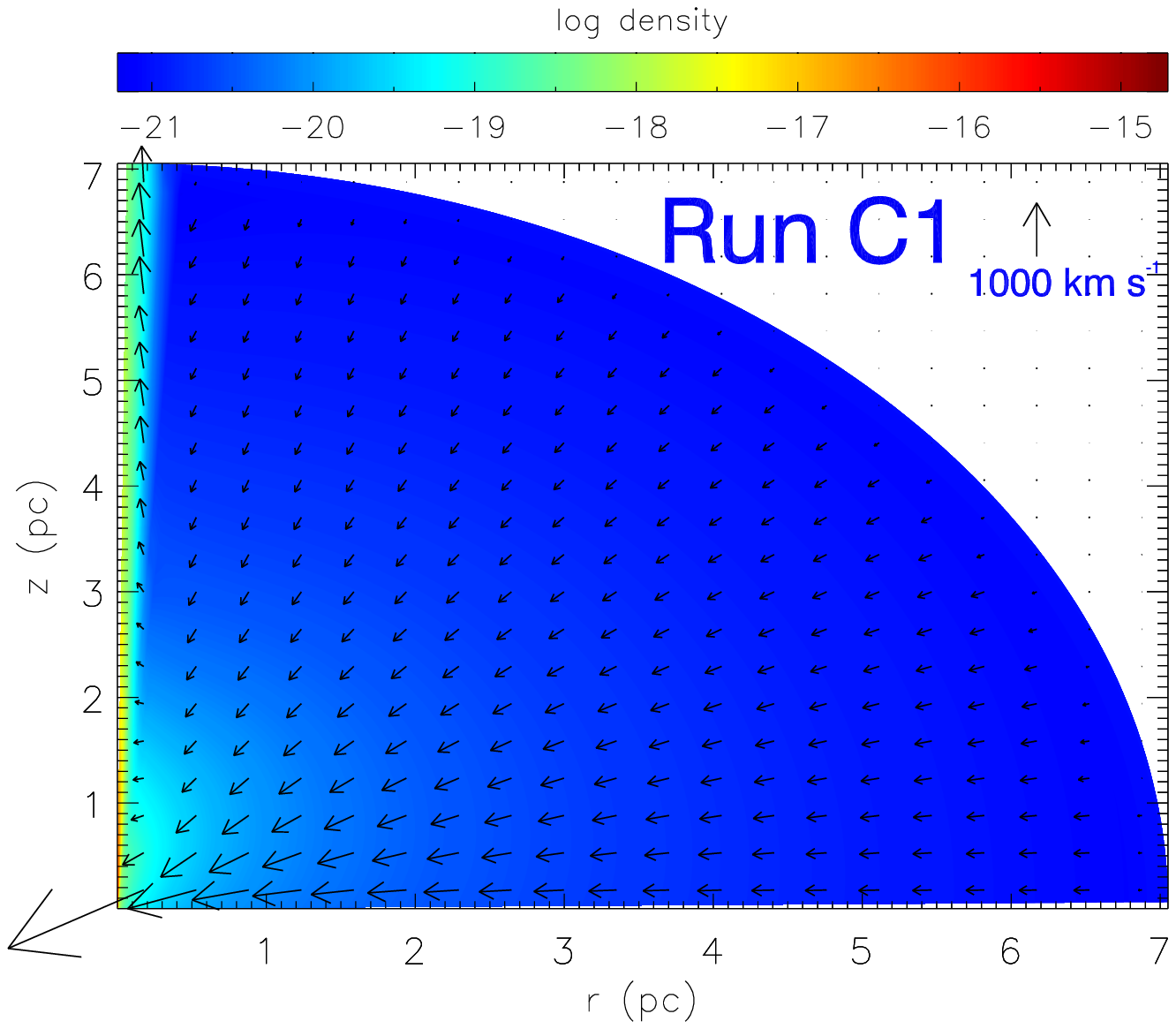}}}
\scalebox{0.29}[0.35]{\rotatebox{0}{\includegraphics[bb=90 340 480 700]{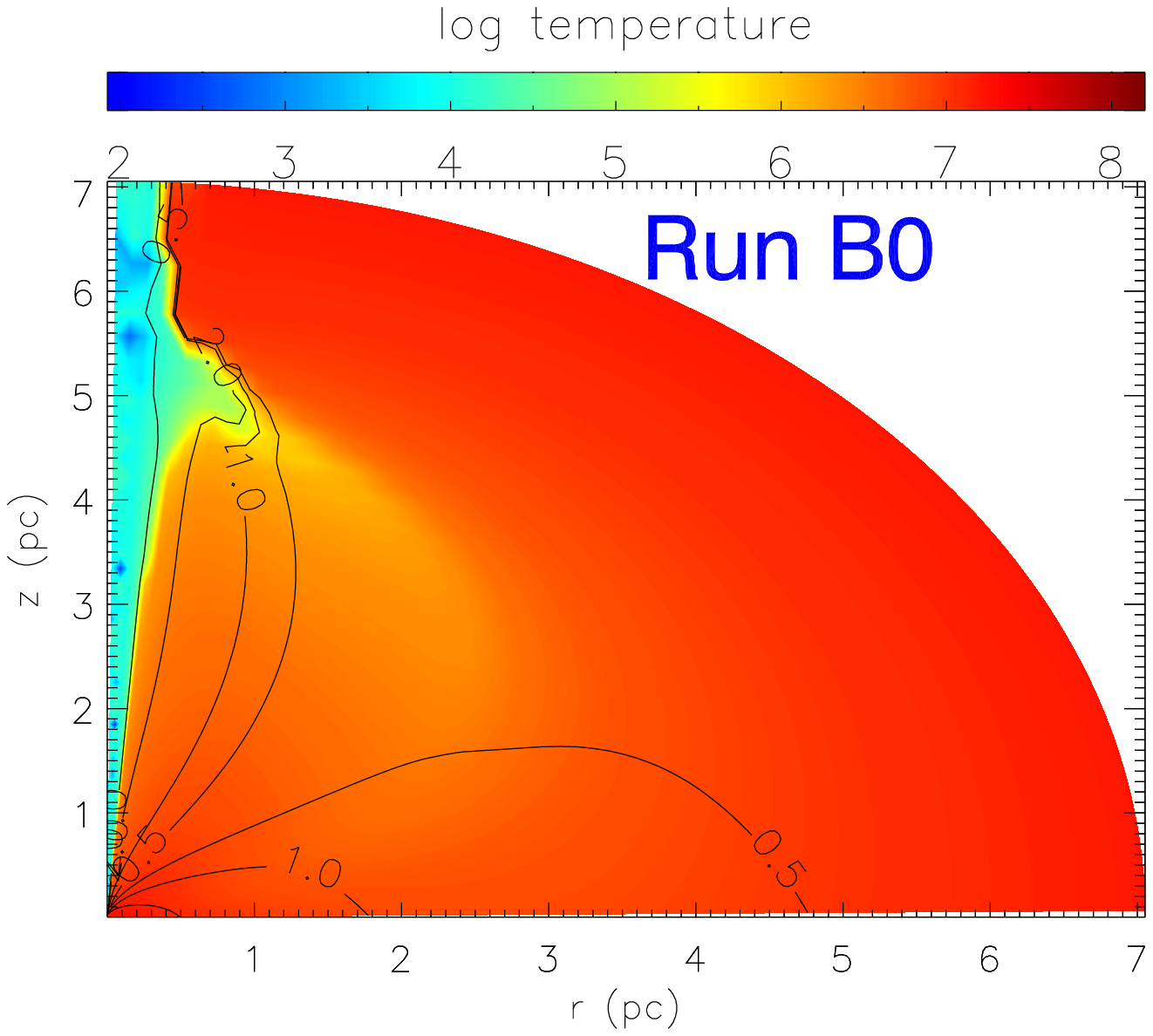}}}
\scalebox{0.29}[0.35]{\rotatebox{0}{\includegraphics[bb=70 340 480 700]{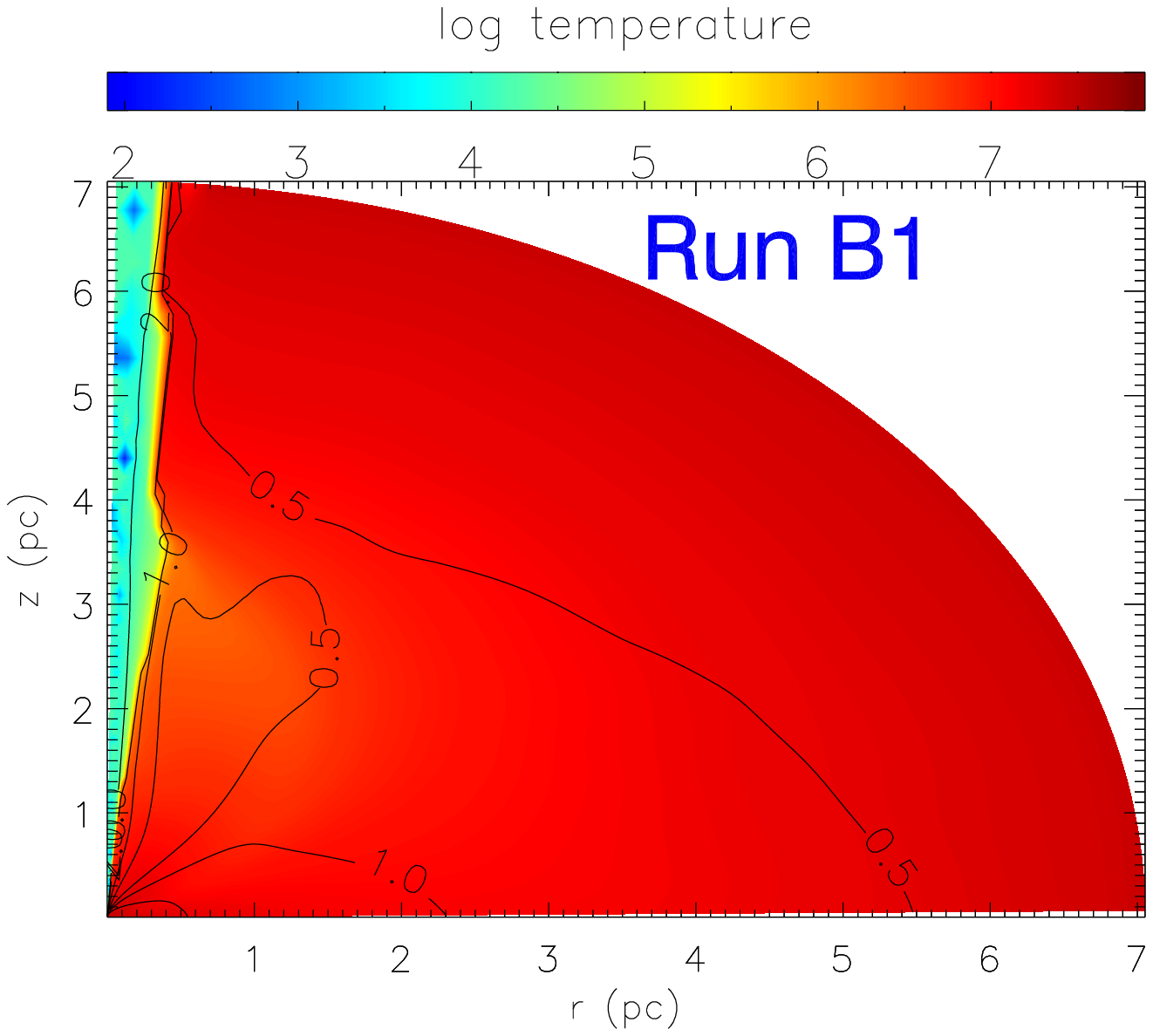}}}
\scalebox{0.29}[0.35]{\rotatebox{0}{\includegraphics[bb=70 340 480 700]{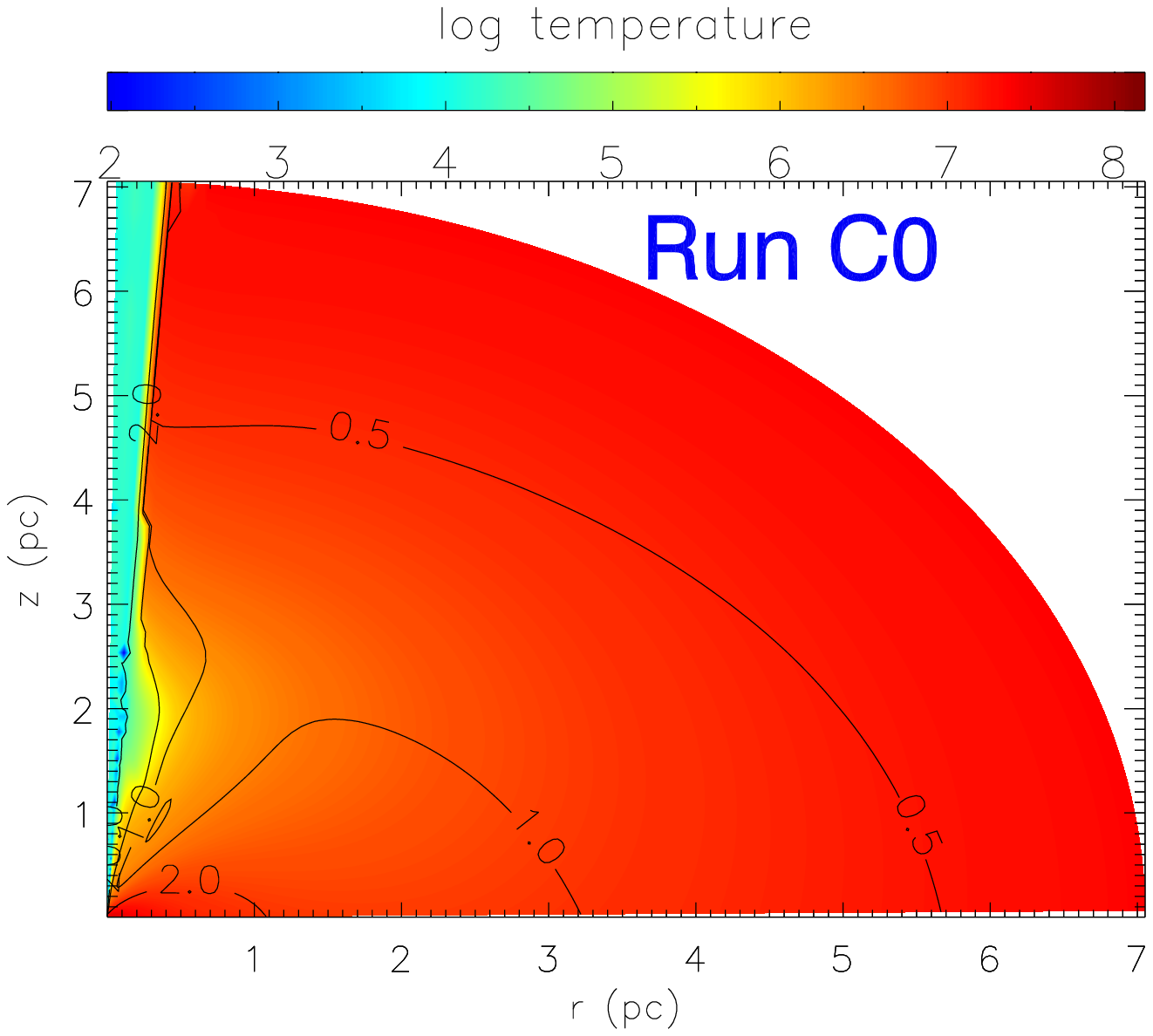}}}
\scalebox{0.29}[0.35]{\rotatebox{0}{\includegraphics[bb=70 340 480 700]{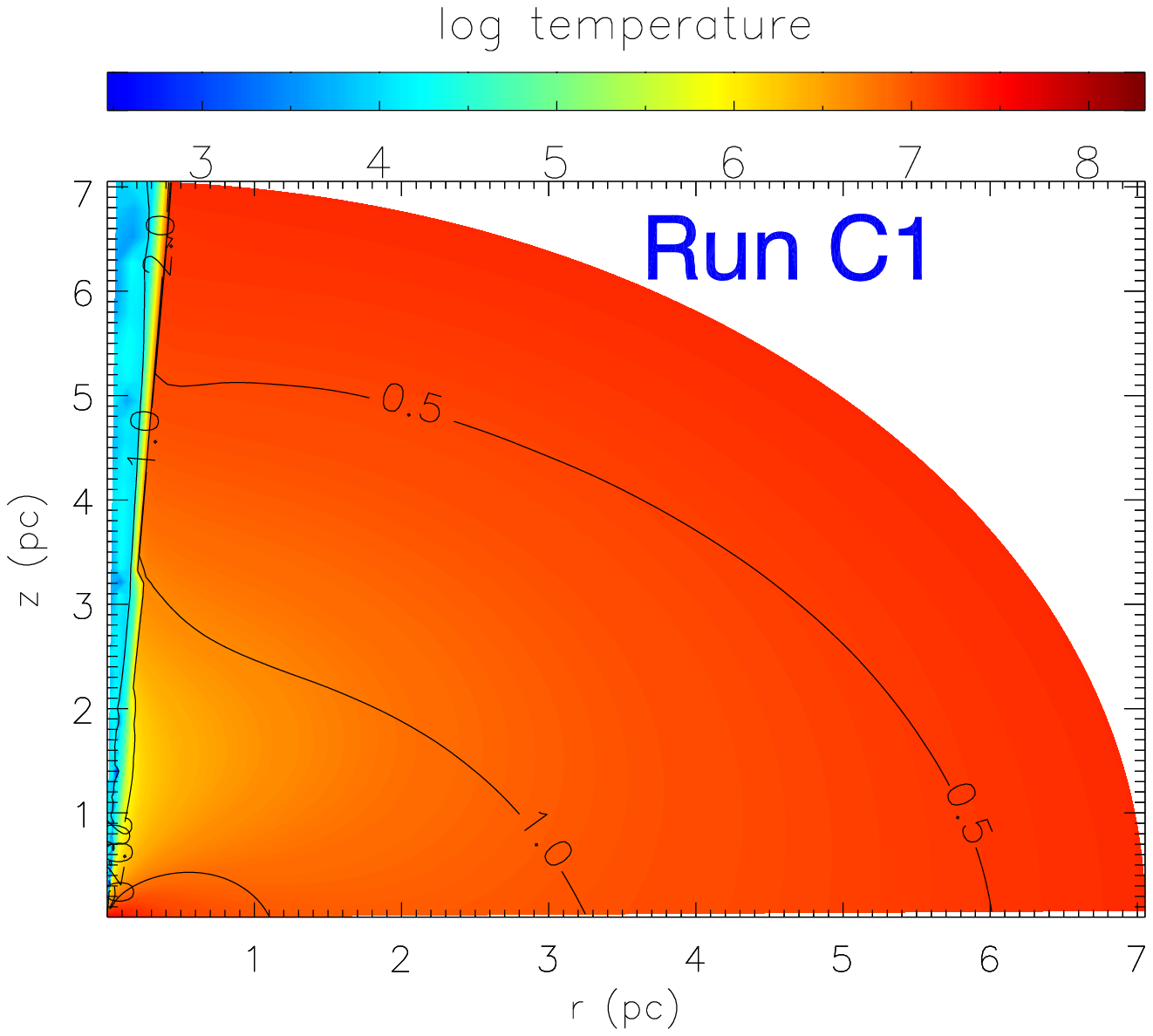}}}
 \ \centering \caption{Snapshot of properties of the irradiated flows at $t=0.8$ (in units of the orbital time at the outer boundary of the computational domain).
 Upper panels: logarithm density (colour) overplotted by the poloidal velocity vecter (arrows); Bottom panels: logarithm temperature (colour) overplotted by contours
 of Mach number.}
 \label{fig 2}
\end{figure*}

\begin{figure*}

\scalebox{0.4}[0.45]{\rotatebox{0}{\includegraphics[bb=60 20 480 360]{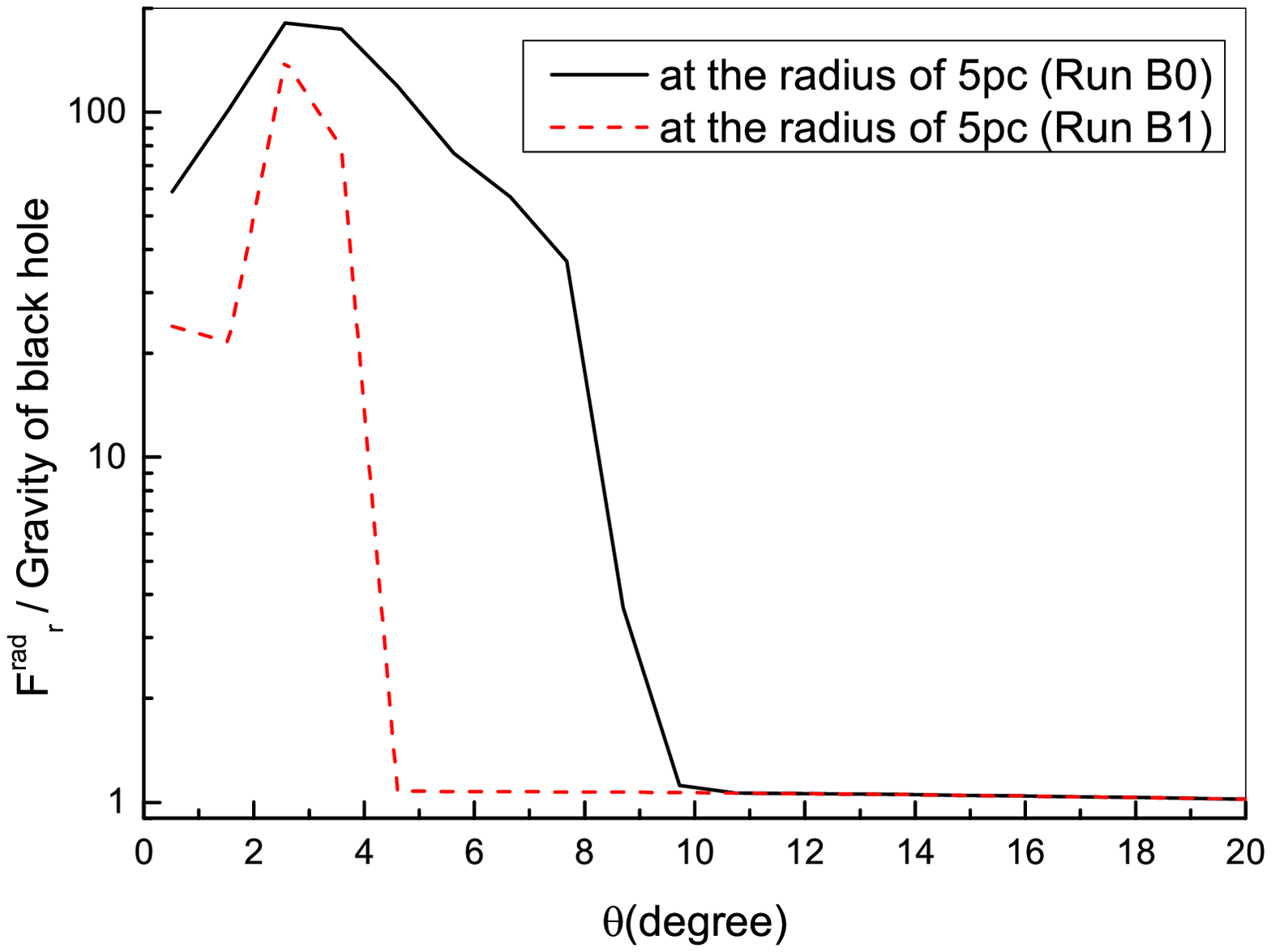}}}
\scalebox{0.4}[0.45]{\rotatebox{0}{\includegraphics[bb=30 20 480 360]{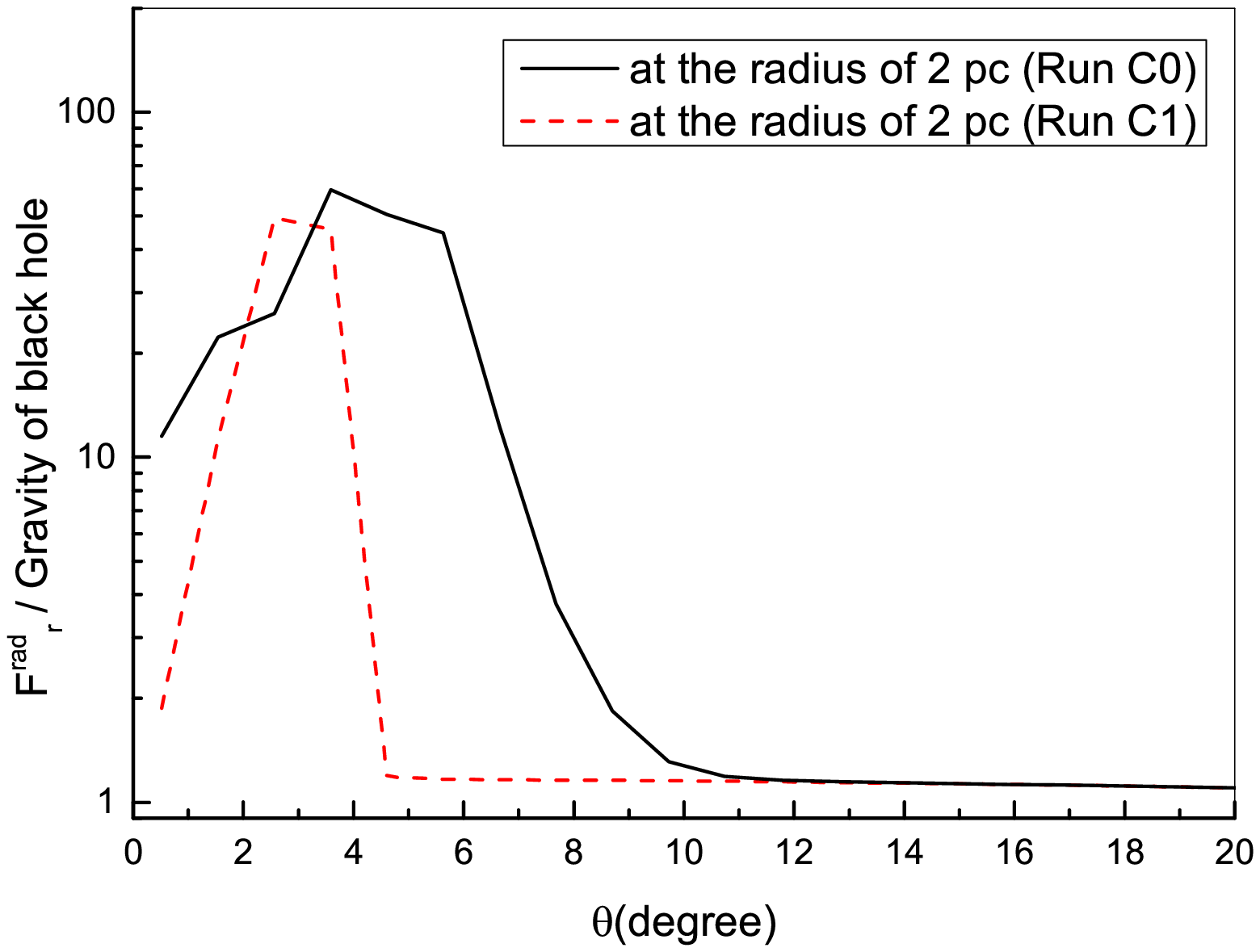}}}

 \ \centering \caption{The time-averaged angle distribution of ratio of the radial component of radiation force to the gravity of black hole.}
 \label{fig 3}
\end{figure*}

\subsection{Model setup}
We set a non-rotating black hole with mass $M_{\rm BH}=10^8\MSUN$ at the origin. The schwarzschild radius of black hole is $r_{\rm s}=2GM_{\rm BH}/c^2$. The accretion disc inner radius ($r_\ast$) is assumed to be at the marginally stable orbit, i.e. $r_\ast=~3~r_{\rm s}~=~8.8~\times~10^{13}$~cm. The computational domain is located in the angular range of $0 \leq \theta \leq \pi/2$ and the radial range of $500 r_{*} \leq r \leq 2.5\times10^5 r_{*}$. In the radial direction, we divide the radial range into 144 non-uniform zones in which the zone size radio is fixed at $(\bigtriangleup r)_{i+1} / (\bigtriangleup r)_{i} = 1.04$. In the angular direction, we divide the polar angle range into 88 uniform zones, i.e. $(\bigtriangleup \theta)_{j+1} / (\bigtriangleup \theta)_{j} = 1.0$.

Initially, the computational domain is uniformly filled with a rarefied gas, whose density and temperature are assumed to be $\rho_0$ and $T_0$, respectively. The initial rotational velocity of the gas is assumed to be
\begin{equation}
v_\phi(r, \theta)=l_0 (1-|cos{\theta}|)/(r \sin{\theta}).
\end{equation}
Here, the specific angular momentum on the equatorial plane is expressed by $l_0=c r_{\ast}\sqrt{r'_{\rm c}/6}$, where $r'_{\rm c}$ is the ``circularization radius'' on the equatorial plane in units of $r_{\ast}$ for the Newtonian potential. When $r'_{\rm c}$ is set to be zero, the gas is non-rotational. The initial velocities in the polar and radial direction are assumed to be zero.

For the polar boundaries, we apply the axially symmetric boundary condition at the pole (i.e. $\theta=0$) and the reflecting boundary condition at the equator (i.e. $\theta=\pi/2$). At the inner radial boundary, we employ the outflow boundary condition, i.e when $v_{r}<0$, the values of physical variables in the ghost zones are set to the values in the corresponding active zones; when $v_{r}>0$, $v_{r}(r,\theta)$ in the ghost zones is set to be zero, whereas the other variable values are set
to the values in the corresponding active zones. At the outer radial boundary, the variable values in the corresponding active zones are copied to the ghost zones when $v_{r}>0$; when $v_{r}<0$, $v_{\theta}(\theta)=v_{r}(\theta)=0$, $v_{\phi}(r,\theta)=v_{\phi}(2.5\times10^5 r_{*},\theta)$, and the density and temperature are set $\rho_0$ and $T_0$, respectively. The outer boundary condition is used to mimic the situation where there is always static gas available for accretion.

\begin{figure}

\scalebox{0.5}[0.5]{\rotatebox{0}{\includegraphics[bb=40 20 500 360]{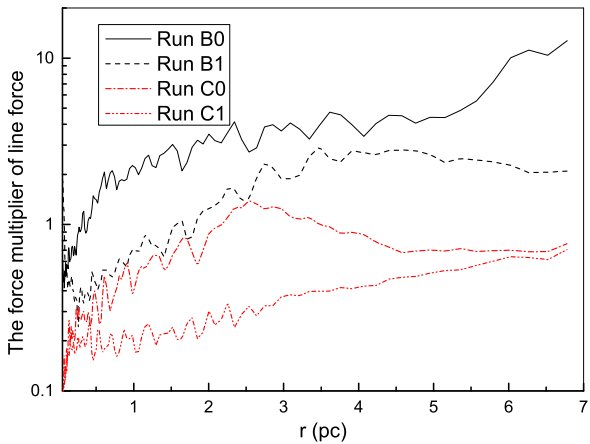}}}

 \ \centering \caption{The time-averaged radial distribution of the averaged force multiplier at the pole.}
 \label{fig 4}
\end{figure}

To calculate the net cooling rate of the X-ray irradiated flows, we assume the characteristic temperature of the X-ray radiation $T_{\rm X}=8\times 10^7$ K (e.g. Proga, Stone, \& Kallman 2000; Sozonov et al. 2005). Therefore, the Compton temperature $T_{\rm C}=0.25 T_{\rm X}=2\times10^7$ K and the corresponding Compton radius, at which the local isothermal sound speed at the Compton temperature is equal to the escape velocity,  $R_{\rm C}=G M_{\rm BH} \mu m_{\rm p}/kT_{\rm C}=9\times10^4r_{\ast}$, where the mean molecular weight $\mu=1$ is set. We note that the Compton radius is inside our computational domain. Finally, we set the absorption coefficient $\kappa_{\rm X}=0.4\text{ g}^{-1}\text{cm}^2$ for all $\xi$ to calculate the attenuation of the X-rays and assume the flows is optically thin in UV bands.

\section{Results}

\subsection{Effect of Nuclear Stars Gravity on the pc-scale non-rotating flows}
Here, we study the role of nuclear stars gravity in dynamics of the irradiated flows by quasar. For simplicity, the flow irradiated by quasar is assumed to be non-rotating and the total accretion luminosity of disc is kept fixed. The accretion luminosity is given by $L=\eta \MDOT_{\rm a} c^2=2\eta G M_{\rm BH}\MDOT_{\rm a}/r_{\rm s}$, where $\eta$ and $\MDOT_{\rm a}$ are the rest-mass conversion efficiency and the mass accretion rate on the BH, respectively. We adopt $\eta=~0.0833$ and set $\MDOT_{\rm a}$ to be $10^{26}~{\rm g~s^{-1}}$, and then the disc yields the accretion luminosity, $L = 0.6 L_{\rm Edd}$, where $L_{\rm Edd}$ is the Eddington luminosity. The corresponding accretion rate of Eddington luminosity is called as Eddington accretion rate $\dot{M}_{\rm Edd}=L_{\rm Edd}/\eta c^2$.

All the models are initially set to a rarefied gas with density $\rho_{0}=1\times10^{-21}\text{g cm}^3$ and gas temperature $T_{0}=2\times10^7 \text{K}$. If $T_{0}=2\times10^7 \text{K}$ is considered as the gas temperature at infinity, the corresponding Bondi radius $R_{\rm B}=5.5\times10^4 r_{\ast}$ (Bondi 1952), which is located within our computational domain. The other properties of models are summarized in Table 1, where columns (2) and (3) give $f_{\rm UV}$ and $f_{\rm X}$, respectively; column (4) gives the circularization radius $r'_{\rm c}$; column (5) gives the star velocity dispersion $\sigma_{\ast}$ which represents the strength of
potential of the star cluster. Some of the gross properties of the solutions are listed in Columns (6)$-$(11); columns (6) and (7) give the mass inflow rate ($\MDOT_{\rm in}$) and the mass outflow rate ($\MDOT_{\rm out}$) through the radius of $6\times10^5 r_{\rm s}$; column (8) gives the net mass flux through the inner boundary ($\MDOT_{\rm net}(r_{\rm in})$); column (9) gives the maximum outflow velocity at the outer boundary ($v_{r}$); columns (10) and (11) give the kinetic energy ($P_{\rm k}$) and the thermal energy ($P_{\rm th}$) carried out by the outflow through the radius of $r=6\times10^5 r_{\rm s}$. In Table 1, $r'_{\rm c}=0$ means that
the gas is non-rotating.

\begin{figure*}

\scalebox{0.40}[0.35]{\rotatebox{0}{\includegraphics[bb=40 20 500 360]{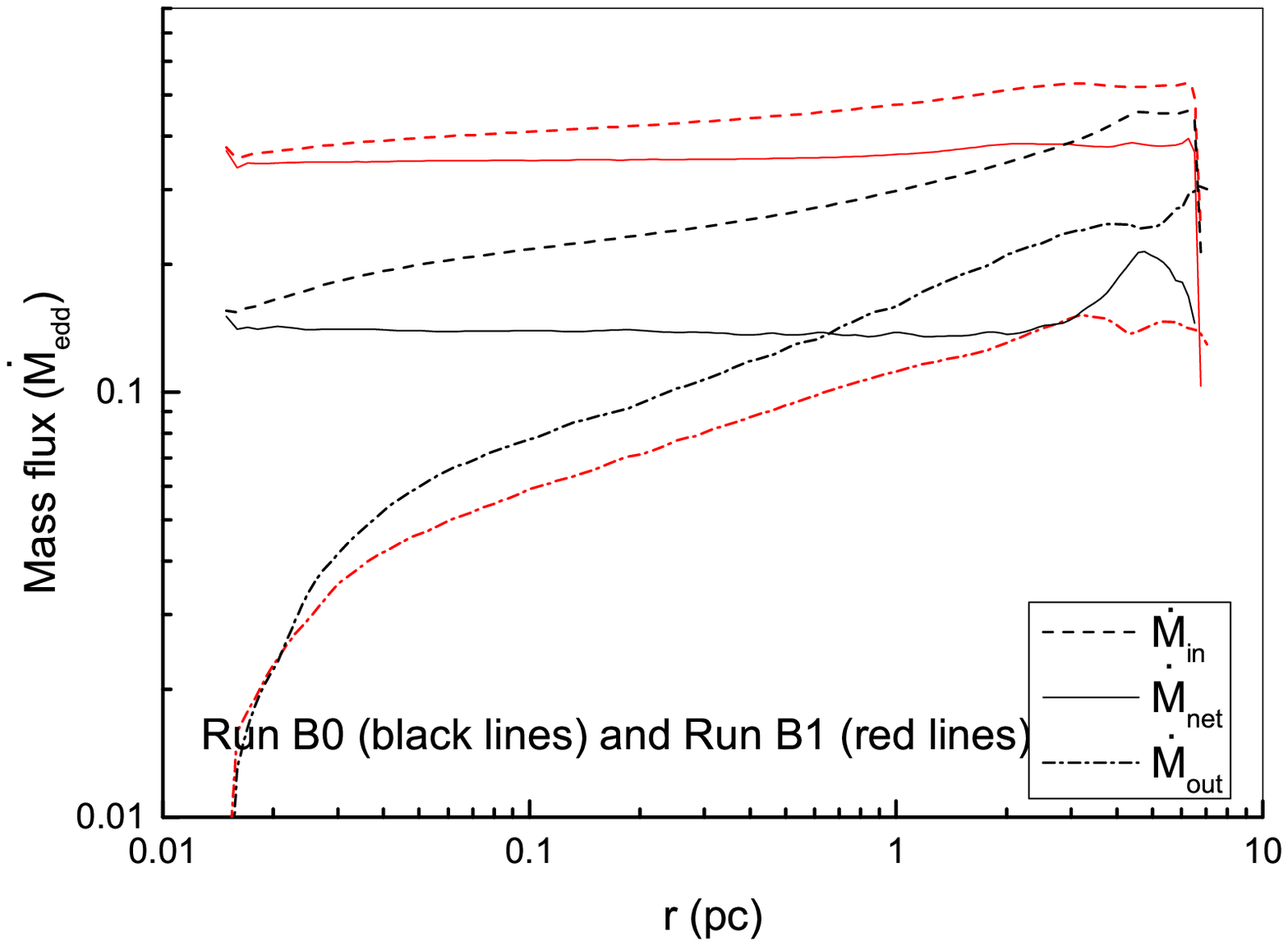}}}
\scalebox{0.40}[0.35]{\rotatebox{0}{\includegraphics[bb=40 20 500 360]{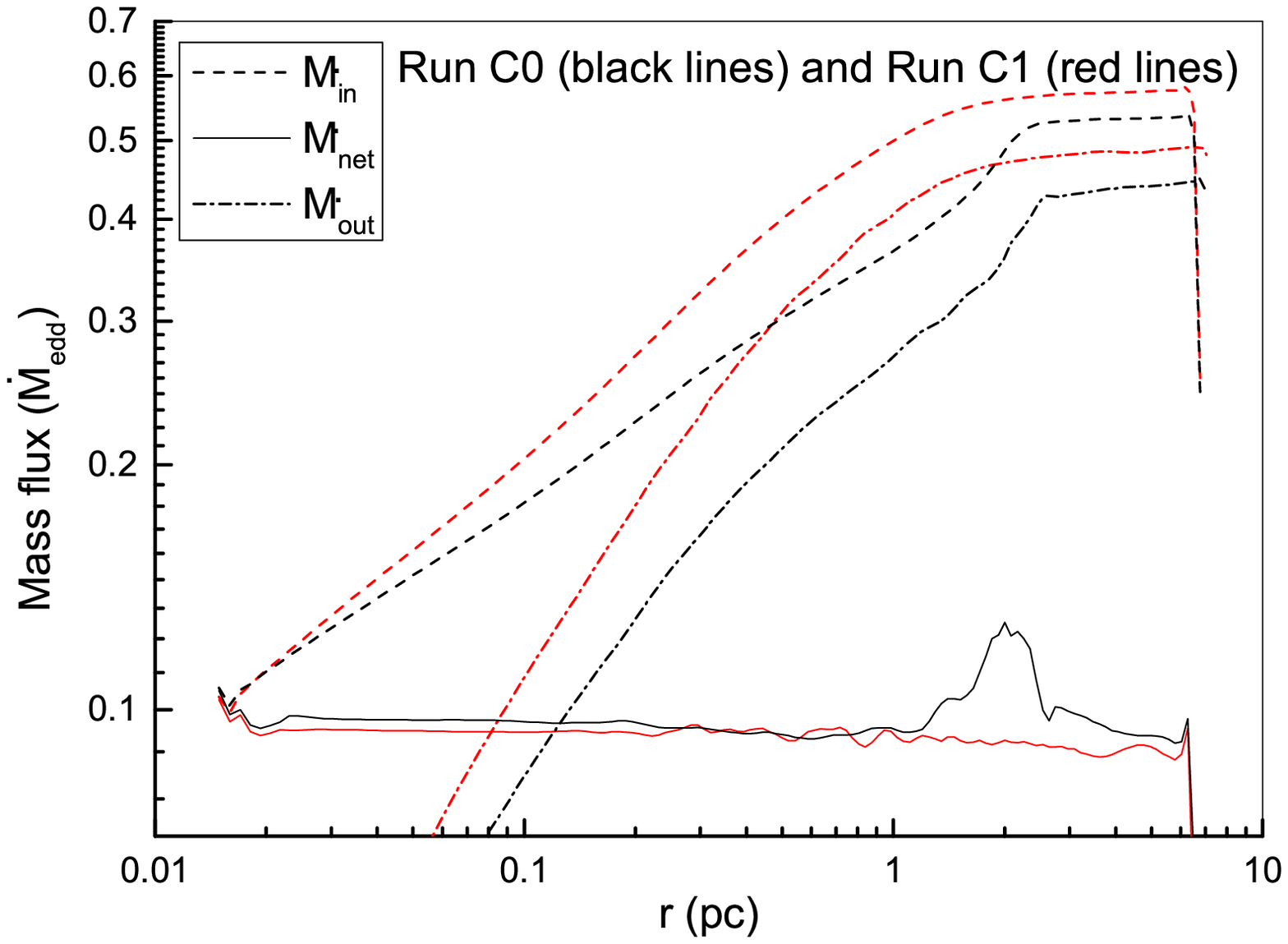}}}
\scalebox{0.40}[0.35]{\rotatebox{0}{\includegraphics[bb=40 20 500 360]{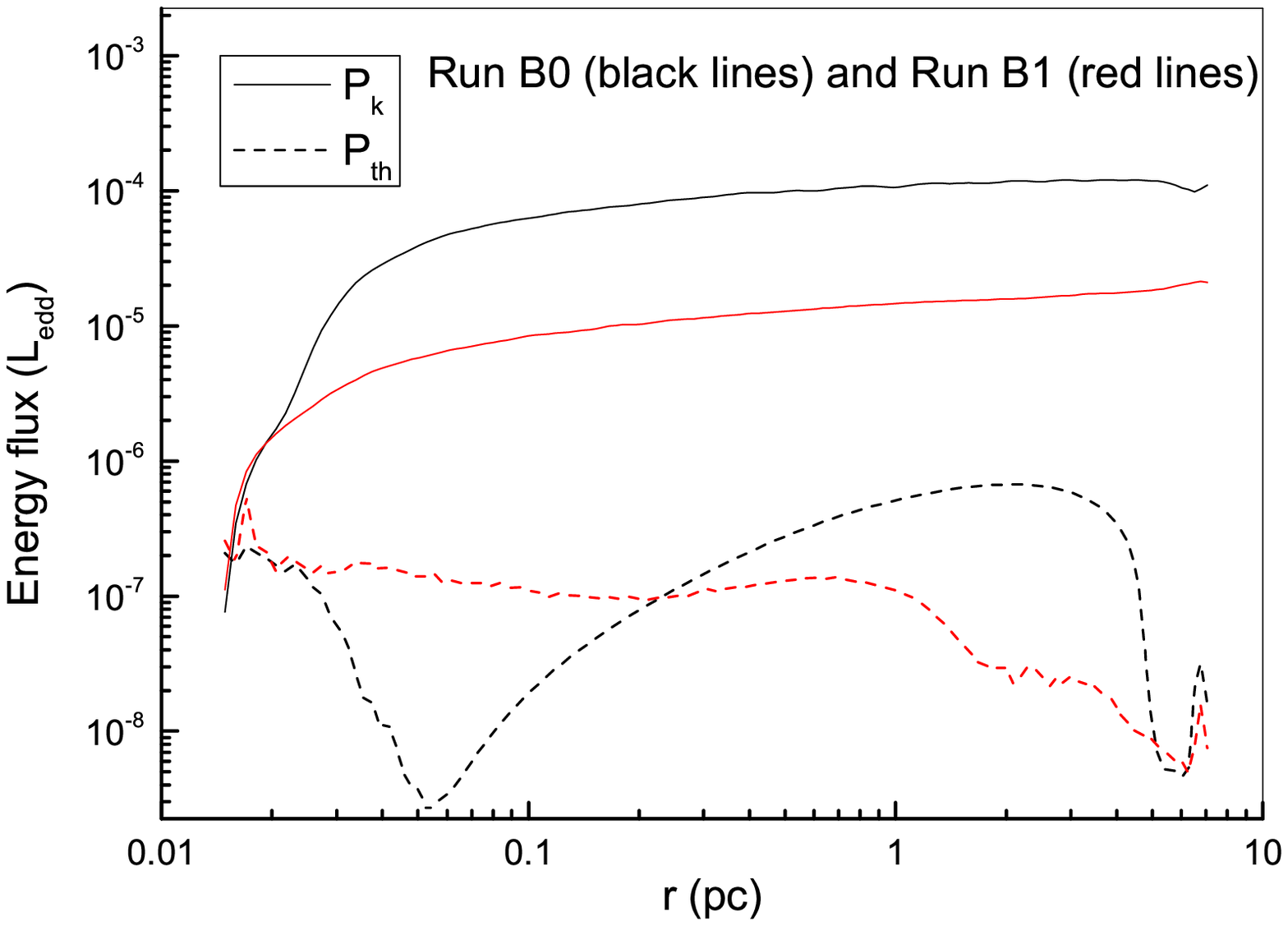}}}
\scalebox{0.40}[0.35]{\rotatebox{0}{\includegraphics[bb=40 20 500 360]{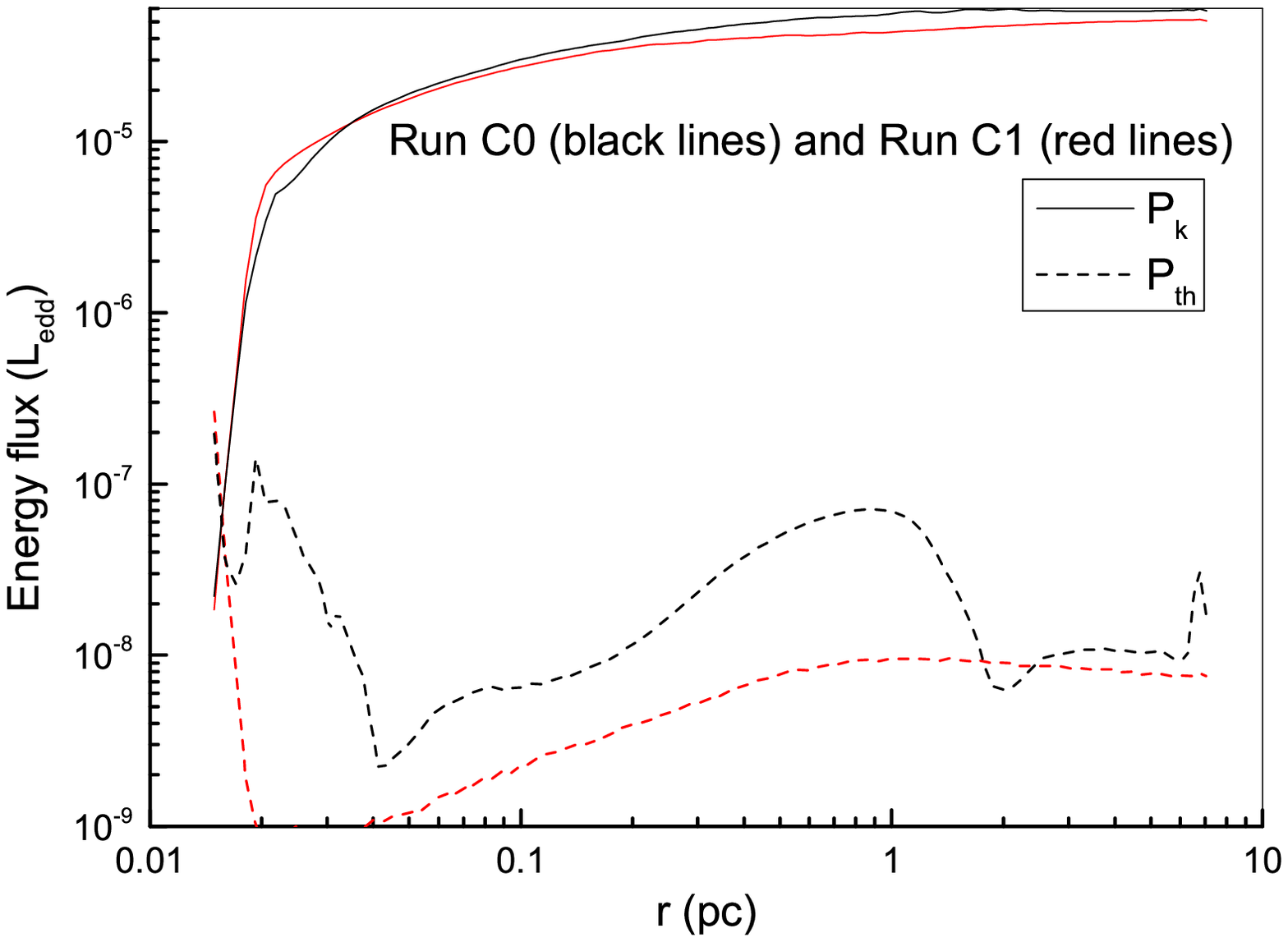}}}
 \ \centering \caption{Top panels: Time-averaged mass fluxes as a function of radius. The solid, dashed, and dotted lines correspond to the net, inflow, and outflow
 rates, respectively. Bottom panels: the energy fluxes carried by the outflow as a function of radius. The solid and dashed lines correspond to the kinetic and thermal
 energy flux, respectively.}
 \label{fig 5}
\end{figure*}

All the models listed in Table 1 reach a steady state after $t=0.3 \text{ }T_{\rm orb}$, where $T_{\rm orb}$ is the orbital period at the outer boundary of the computational domain. As shown in Table 1, in run B1 with nuclear stars gravity, the mass outflow rate is reduced compared to that in run B0. The net accretion rate in run B1 is $\sim$2.5 times that of run B0. In run B0, the terminal maximum velocity of outflow is about half that in run run B0. The kinetic energy carried out by outflow in run B1 is about one fifth of that in run B0. So, the nuclear stars gravity significantly weakens the outflow in models with $f_{\rm X} = 0.2$ and  $f_{\rm UV} = 0.8$. Comparing run C0 and run C1, the mass inflow and outflow rate and the net accretion rate are comparable. The outflow power of run C1 is slightly weaker than that in run C0. In run C1, the terminal maximum velocity of outflow is also slightly reduced by nuclear stars gravity.

\begin{table*}
\begin{center}

\caption[]{Summary of Models in section 3.2}

\begin{tabular}{ccccccccccc}
\hline\noalign{\smallskip} \hline\noalign{\smallskip}

Run &  $f_{\text{\rm UV}}$ & $f_{\rm X}$ &  $r'_{\rm c}$ & $\sigma_{\ast}$   & $\dot{M}_{\text{in}}$($6\times10^5 r_{s}$)  & $\dot{M}_{\text{out}}$($6\times10^5
r_{s}$) & $\dot{M}_{\text{net}}$($r_{\rm in}$) & $v_{\text{r}}$ & $P_{\text{k}}$ & $P_{\text{th}}$ \\
    &      & & ($r_{\ast}$) & ($100\text{ km s}^{-1}$) & ($\dot{M}_{\rm Edd}$) & ($\dot{M}_{\rm Edd}$) & ($\dot{M}_{\rm Edd}$) & ($\text{km s}^{-1}$) & ($L_{\rm Edd}$)
    & ($L_{\rm Edd}$) \\
(1) & (2)             & (3)                         &  (4)      &     (5) & {6}  & (7) & (8) & (9) & (10)   & (11)     \\

\hline\noalign{\smallskip}
Br0  & 0.8  & 0.2     &300  & 0    &-0.44   &0.12  &-0.29 &1500 &2.6$\times10^{-6}$ &6.6$\times10^{-7}$  \\
Br1  & 0.8  & 0.2     &300  & 2.0  &-0.52   &0.04 &-0.46 &1500 &1.3$\times10^{-6}$ &1.6$\times10^{-7}$ \\

\hline\noalign{\smallskip}
Cr0  & 0.95 &0.05     &300  & 0    &-0.53   &0.22  &-0.27 &1400 &2.4$\times10^{-6}$ &9.7$\times10^{-8}$ \\
Cr1  & 0.95 &0.05     &300  & 2.0  &-0.57   &0.26  &-0.30 &1200 &1.2$\times10^{-6}$ &8.6$\times10^{-8}$ \\

 \hline\noalign{\smallskip} \hline\noalign{\smallskip}
\end{tabular}
\end{center}

\begin{list}{}
\item\scriptsize{\textit{Note}. The mean of columns is the same as in Table 1.}
\end{list}
\label{tab2}
\end{table*}

\begin{figure*}

\scalebox{0.29}[0.35]{\rotatebox{0}{\includegraphics[bb=70 340 480 700]{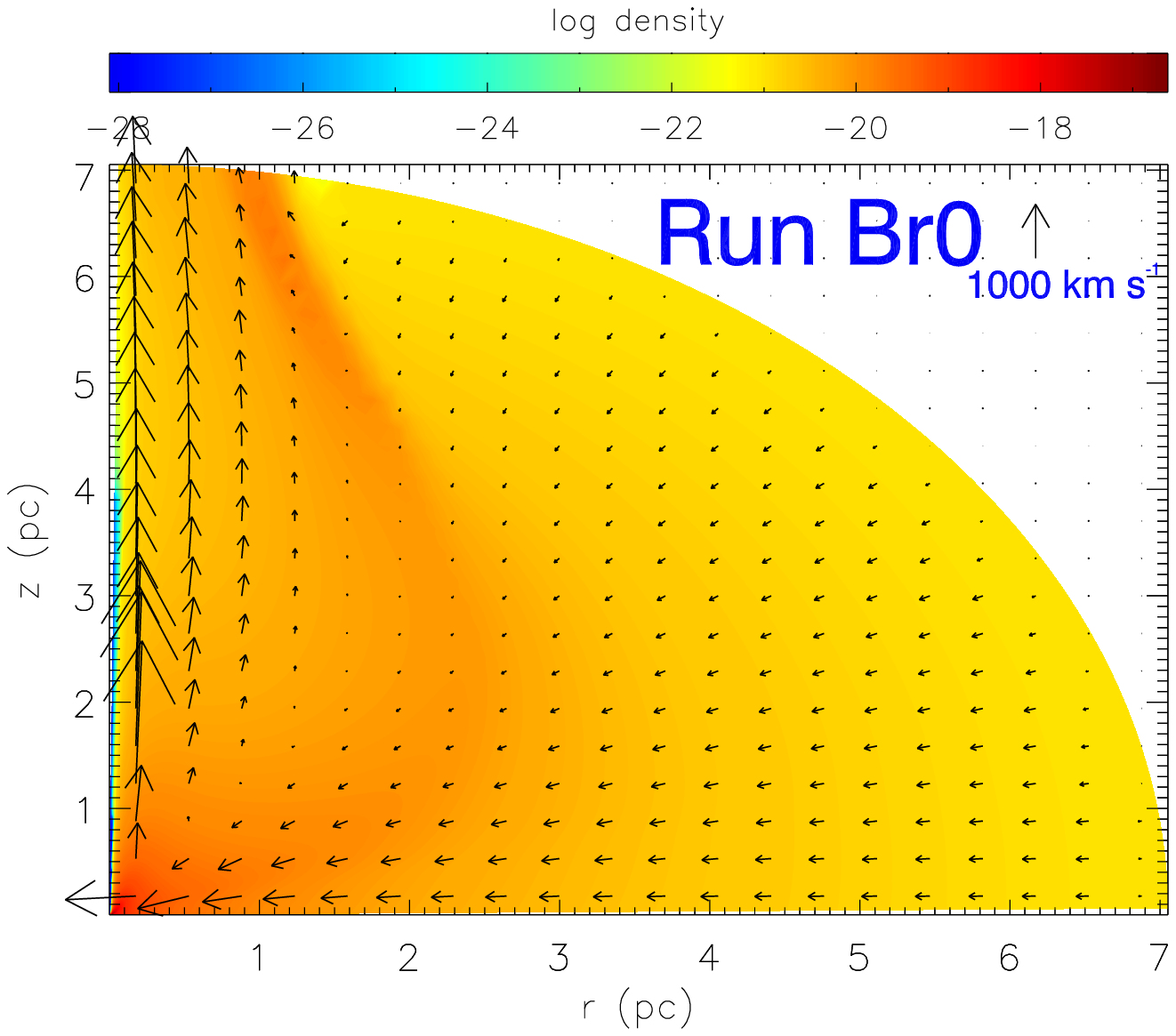}}}
\scalebox{0.29}[0.35]{\rotatebox{0}{\includegraphics[bb=70 340 480 700]{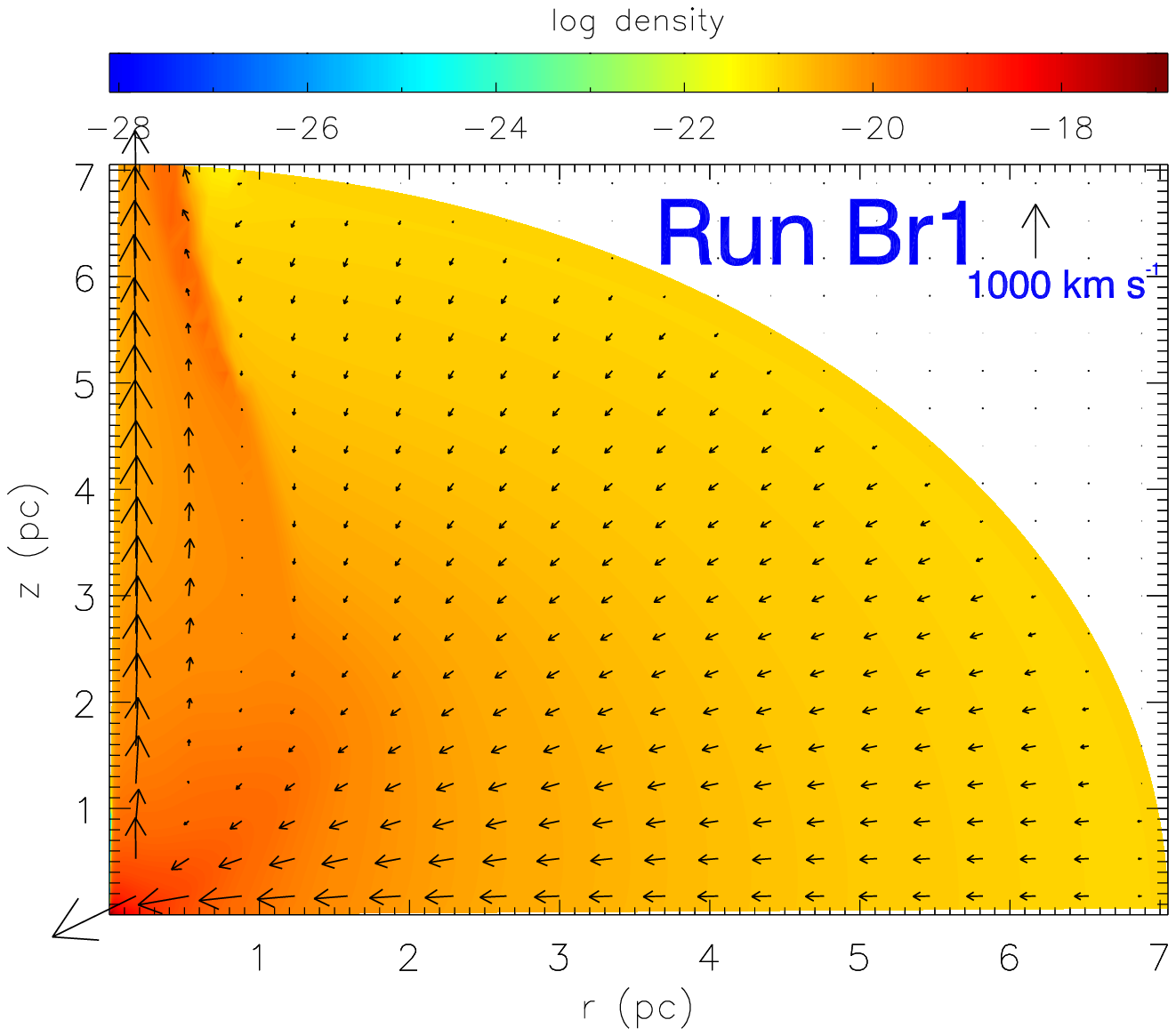}}}
\scalebox{0.29}[0.35]{\rotatebox{0}{\includegraphics[bb=70 340 480 700]{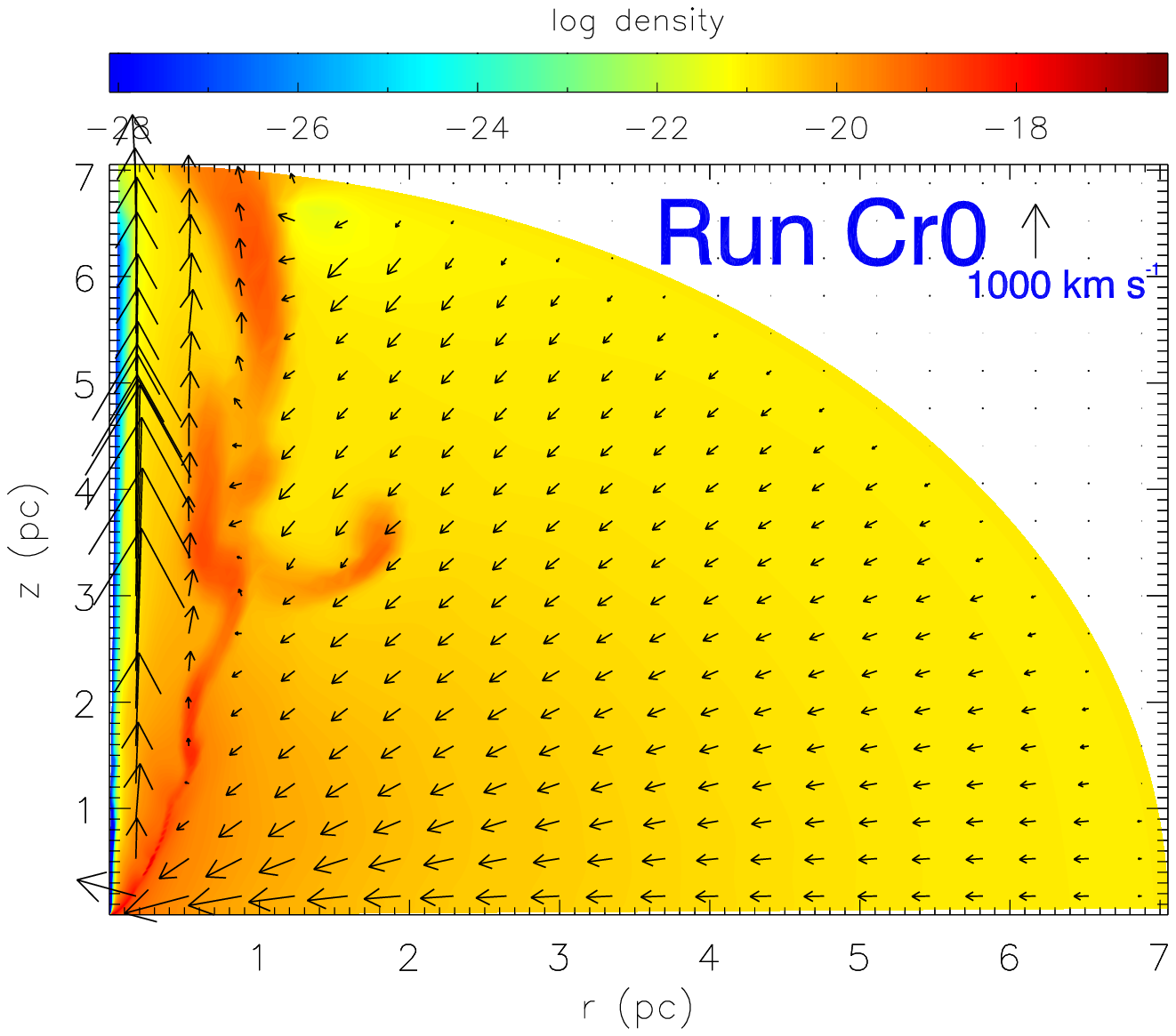}}}
\scalebox{0.29}[0.35]{\rotatebox{0}{\includegraphics[bb=70 340 480 700]{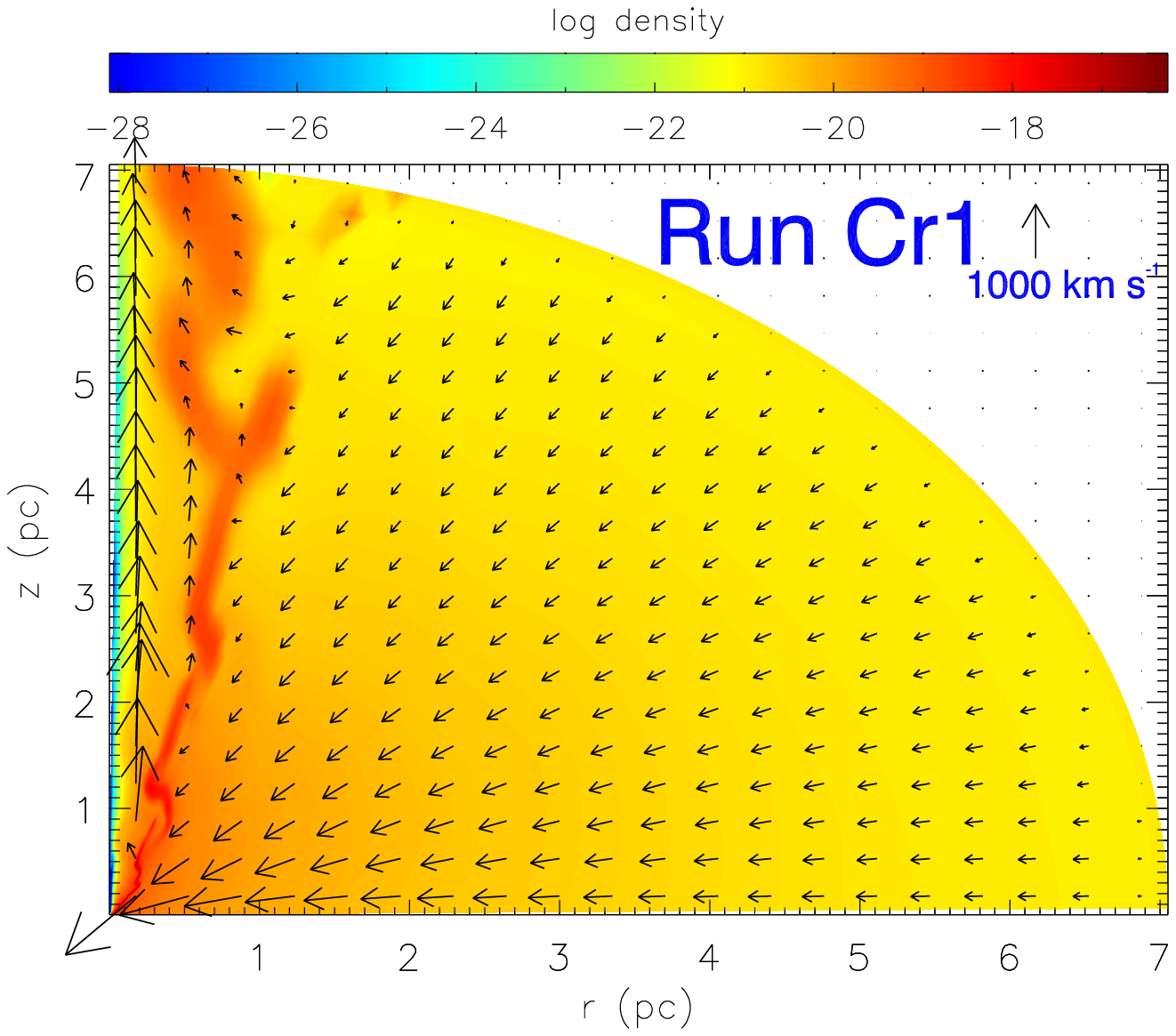}}}
\scalebox{0.29}[0.35]{\rotatebox{0}{\includegraphics[bb=90 340 480 700]{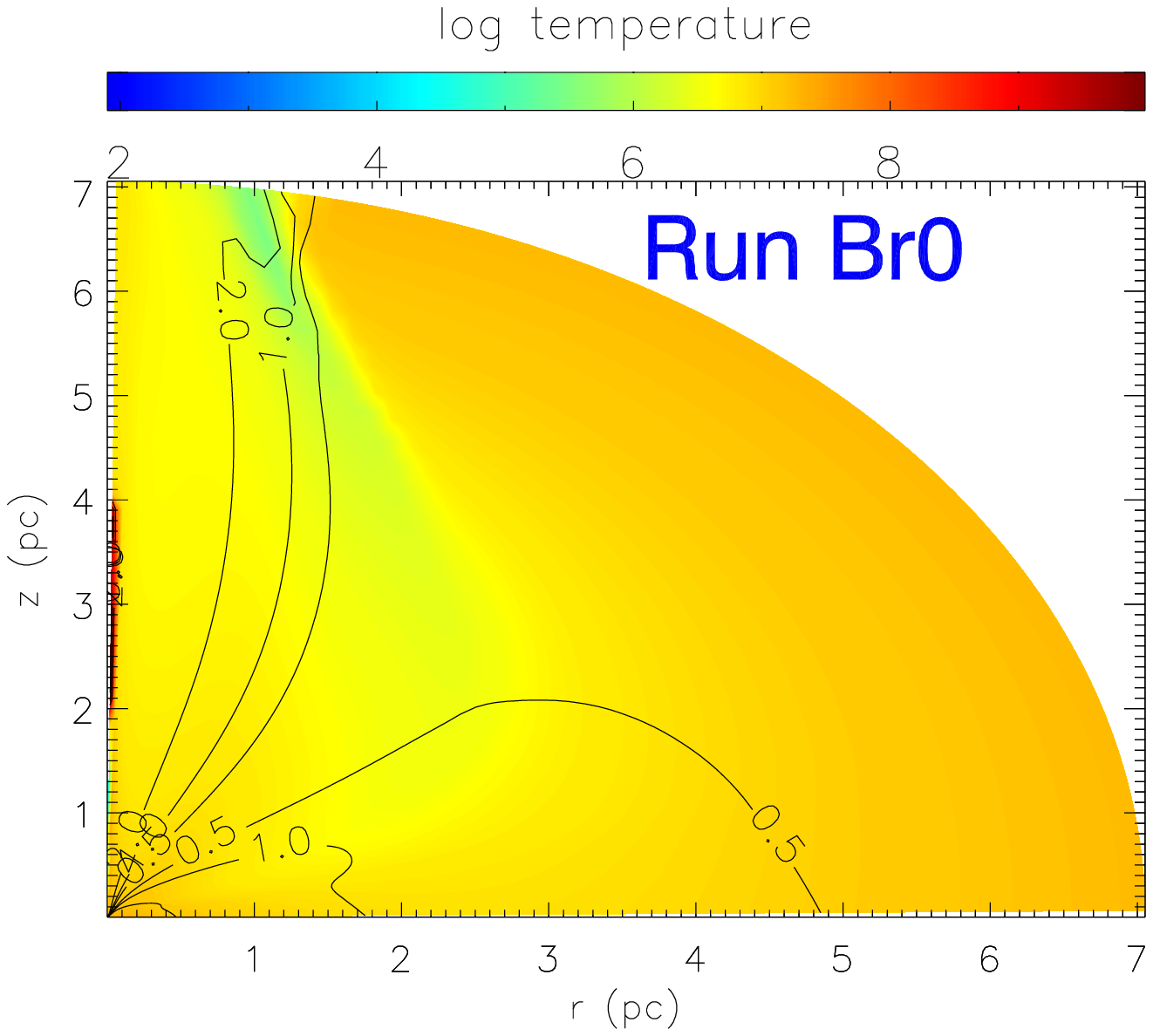}}}
\scalebox{0.29}[0.35]{\rotatebox{0}{\includegraphics[bb=70 340 480 700]{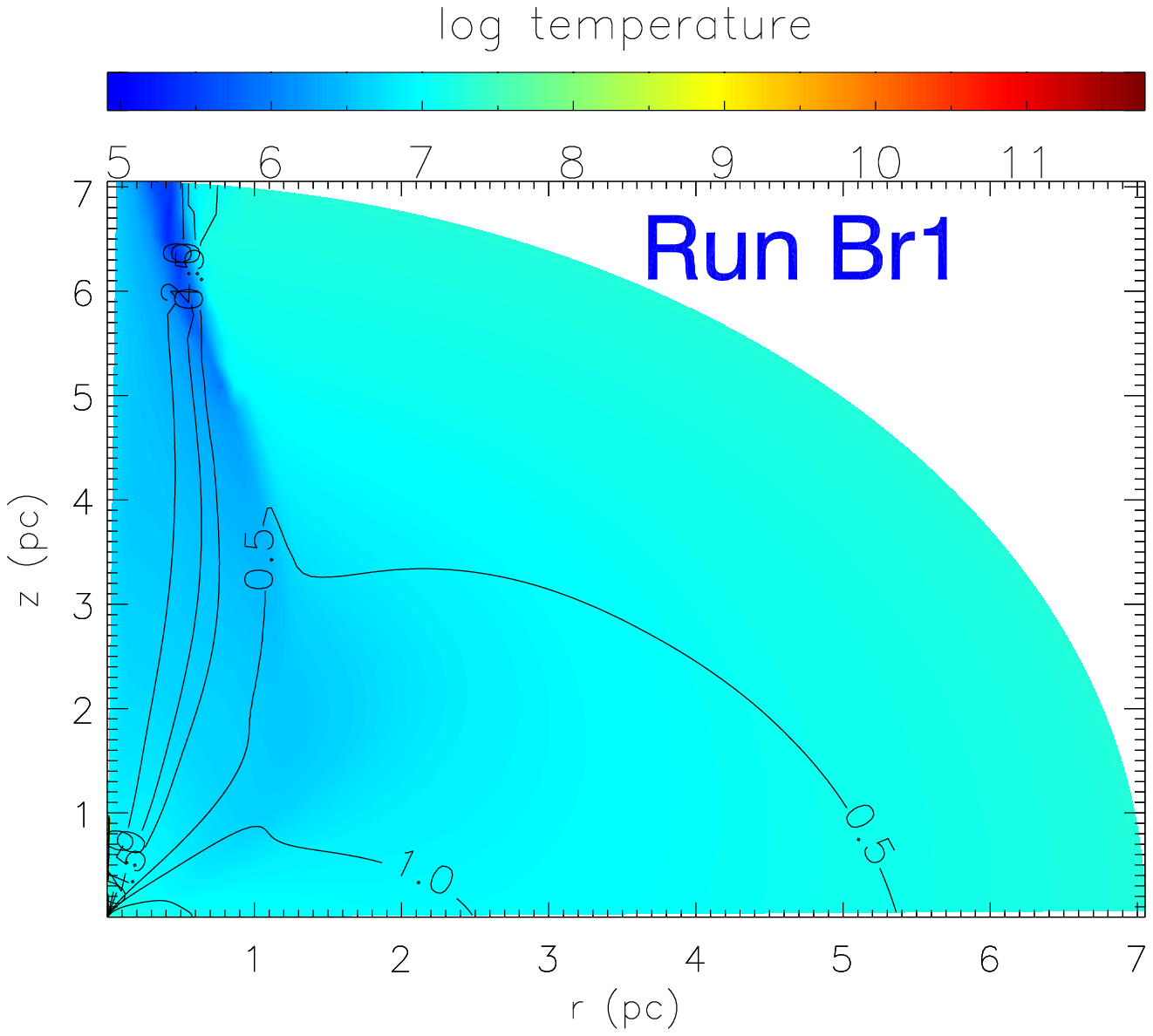}}}
\scalebox{0.29}[0.35]{\rotatebox{0}{\includegraphics[bb=70 340 480 700]{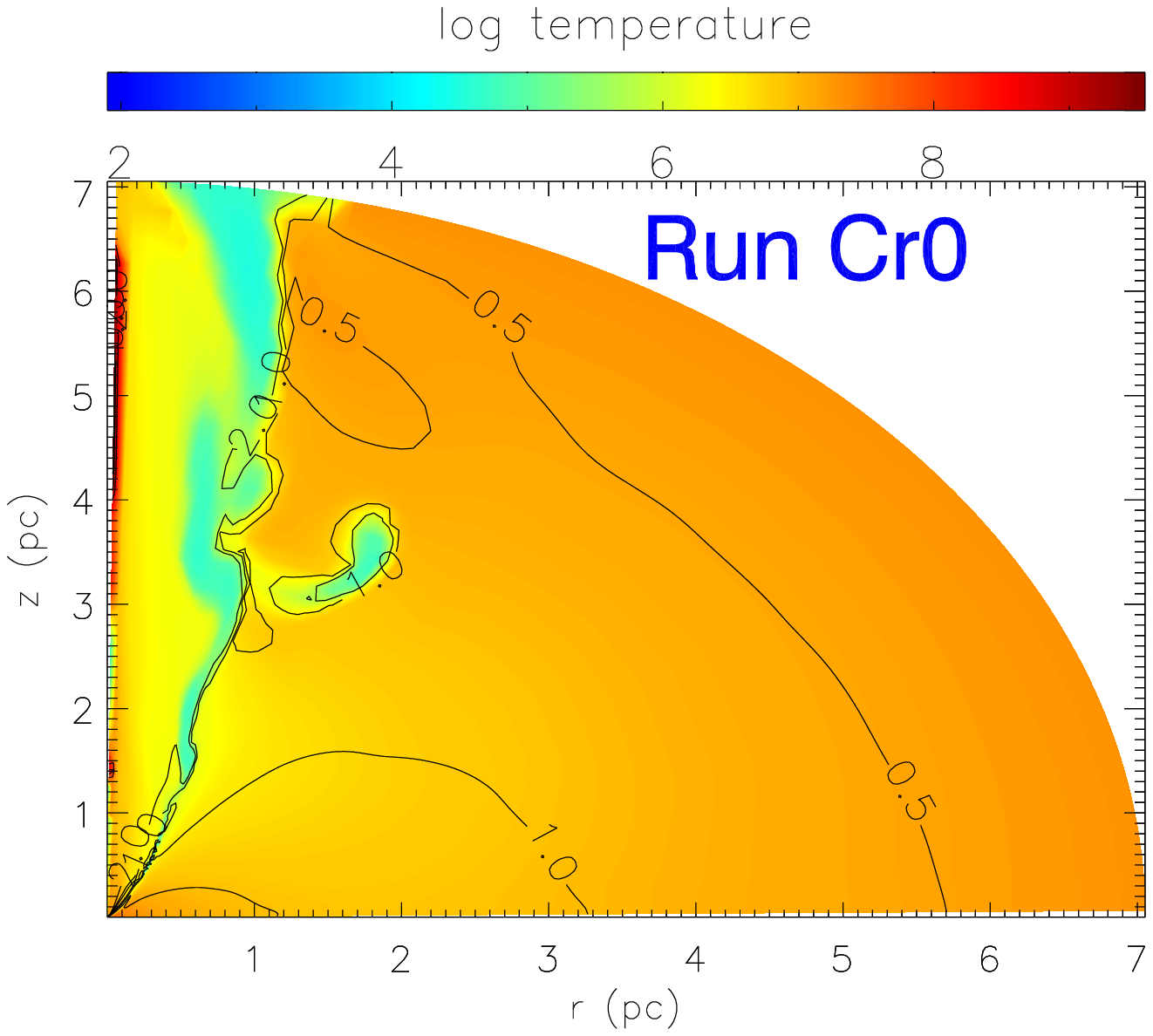}}}
\scalebox{0.29}[0.35]{\rotatebox{0}{\includegraphics[bb=70 340 480 700]{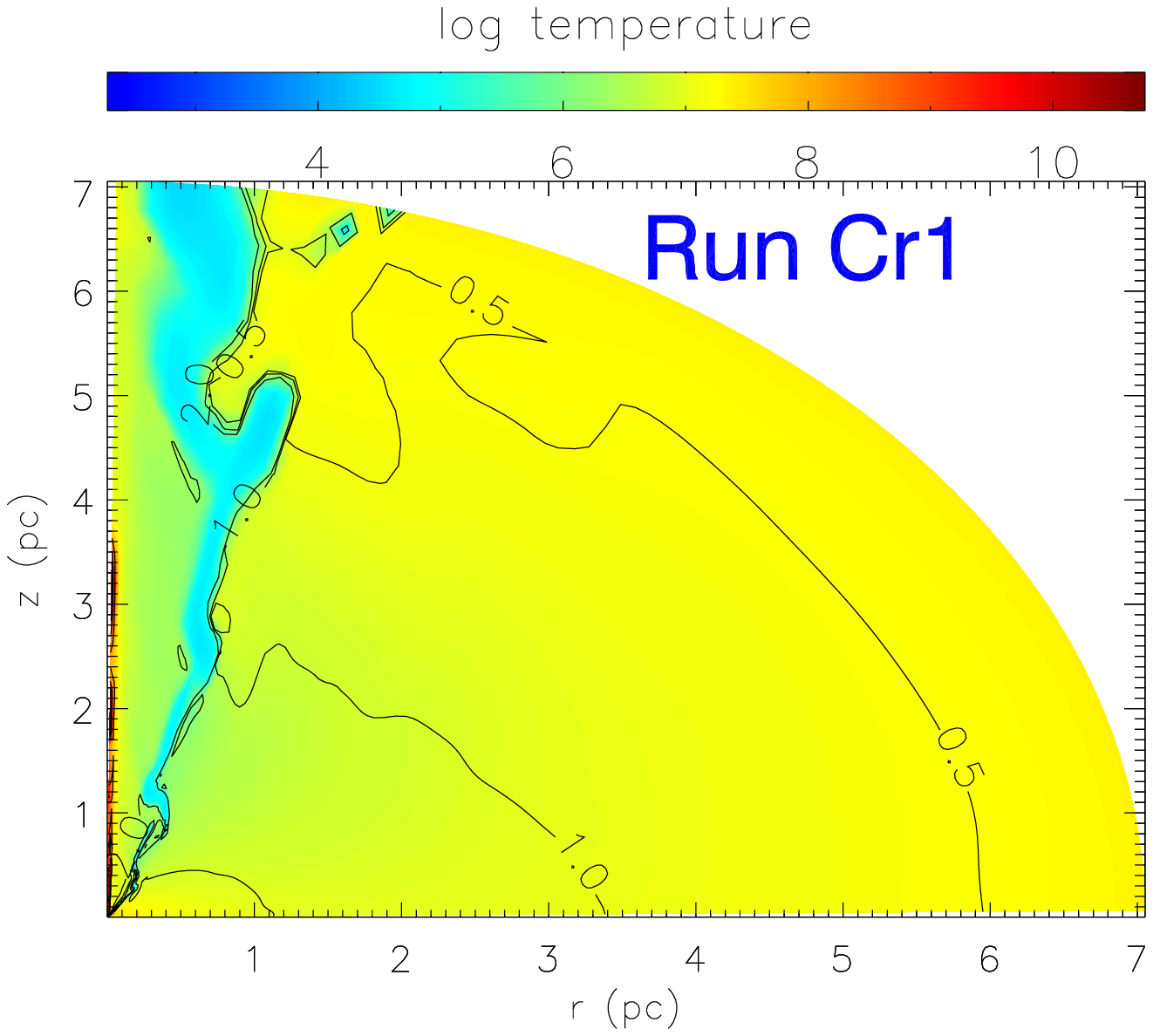}}}

 \ \centering \caption{As in Figure 2, but for runs Br0, Br1, Cr0, and Cr1.}
 \label{fig 6}
\end{figure*}

Figure \ref{fig 2} shows snapshot of properties of the irradiated flows at $t=0.8 \text{ }T_{\rm orb}$. As shown in Figure \ref{fig 2}, nuclear stars gravity changes outflow region in the model of $f_{\rm X} = 0.2$ and  $f_{\rm UV} = 0.8$. For run B0 without nuclear stars gravity, the outflow is strongly pushed toward the pole at $r=\sim5\text{ pc}$; the outflow at the pole is accelerated by line force to the velocity up to $\sim3200\text{ km s}^{-1}$. For run B1 with nuclear star gravity, the starting position of the completely collimated outflow is located at $r=\sim3\text{ pc}$; the terminal velocity of outflow is significantly reduced to $\sim1500 \text{ km s}^{-1}$; the sonic surface of outflow is closer to the pole compared to run B0. In runs C0 and C1, outflow region is much smaller than that in runs B0 and B1 and limited to the pole region.

To study the collimation of outflow, we plot Figure \ref{fig 3},  which shows the time-averaged angle distribution ($0^{o}$--$20^{o}$ away from the pole) of ratio of the radial component of radiation force to gravity. For all runs, when $\theta>\sim 30^{o}$ ($\theta$ is angle away from the pole) the gravitational force of black hole is larger than the radial component of radiation force ($F^{\rm rad}_{r}$). We take the forces at 5 pc as an example to explain the collimation of outflow by stars gravity in the X-ray strong cases (runs B0 and B1). Around the pole, $F^{\rm rad}_{r}$ is much larger than the gravitational force, so that $F^{\rm rad}_{r}$ drives the outflow around the pole. Comparing runs B0 and run B1, $F^{\rm rad}_{r}$ at 5 pc in run B0 is much larger than that in run B1 within $\sim 5^{o}$--$10^{o}$. After considering the stars gravity, the range of $\theta$ angles at which radiation force is larger than gravitational force becomes smaller. Therefore, the angle distribution of outflow in run B1 is much narrower than that in run B0 at 5 pc. The outflow is collimated by the stars gravity. The collimation of outflow in X-ray weak cases happens at small radii ($\sim 2$ pc). The right panel of Figure \ref{fig 3} shows the ratio of radiation force to gravitational force at 2 pc. As in the X-ray strong cases, after considering the stars gravity, the range of $\theta$ angles at which radiation force is larger than gravitational force becomes smaller. The outflow is collimated.

The temperature of outflow is lower than $10^5$ K and is much lower than the temperature of inflow. The outflow is driven by line force produced by UV photons. The line force will become negligible when the temperature of the accreted gas is higher than $10^5$ K. If the inflow gas cools to temperature below $10^5$ K because of radiation losses, the line force becomes significant and drives the inflow gas to be the outflow along the pole. Nuclear stars gravity has two important effects: (1) Nuclear stars gravity increases specific momentum (or radial infall velocity) of the inflow gas, reduces the travel time of the inflow gas from the outer boundary to inner region, and then reduces the radiation losses of the inflow gas. This causes that the inflow gas in runs with stars gravity has slightly higher temperature compared with the case without nuclear stars gravity, which is not helpful to produce larger line force to turn the inflow into the outflow, because the force multiplier $\mathcal{M}$ will become smaller when temperature is increased (see equation 17 in Proga (2007)). As a result, the outflow region contracts toward the centre (or the pole) in the case with nuclear stars gravity, i.e. the outflow is collimated at smaller radii compared with the case without nuclear stars gravity.  (2) Nuclear stars gravity also indirectly reduce the acceleration of the outflow.

Figure \ref{fig 4} shows the radial distribution of averaged $\mathcal{M}$ at the pole (i.e. outflow region). As shown in Figure \ref{fig 4}, the force multiplier is less in run B1 than that in run B0, and even at the outer boundary the $\mathcal{M}$ value of run B1 is about one-sixth of that of run B0, so that the radiation force produced by UV photons is weaker in run B1 than that in run B0. Comparing run C0 and run C1, their $\mathcal{M}$ values are comparable at the outer boundary. In addition, it is noted that the terminal velocity of outflow is smaller in run C0 than that in run B0. From Figure \ref{fig 4}, we note that $\mathcal{M}$ of run C0 is much smaller than that of run B0, and then the line force is weaker in run C0 than that in run B0 though run C0 has higher UV luminosity.

\begin{figure*}

\scalebox{0.40}[0.35]{\rotatebox{0}{\includegraphics[bb=40 20 500 360]{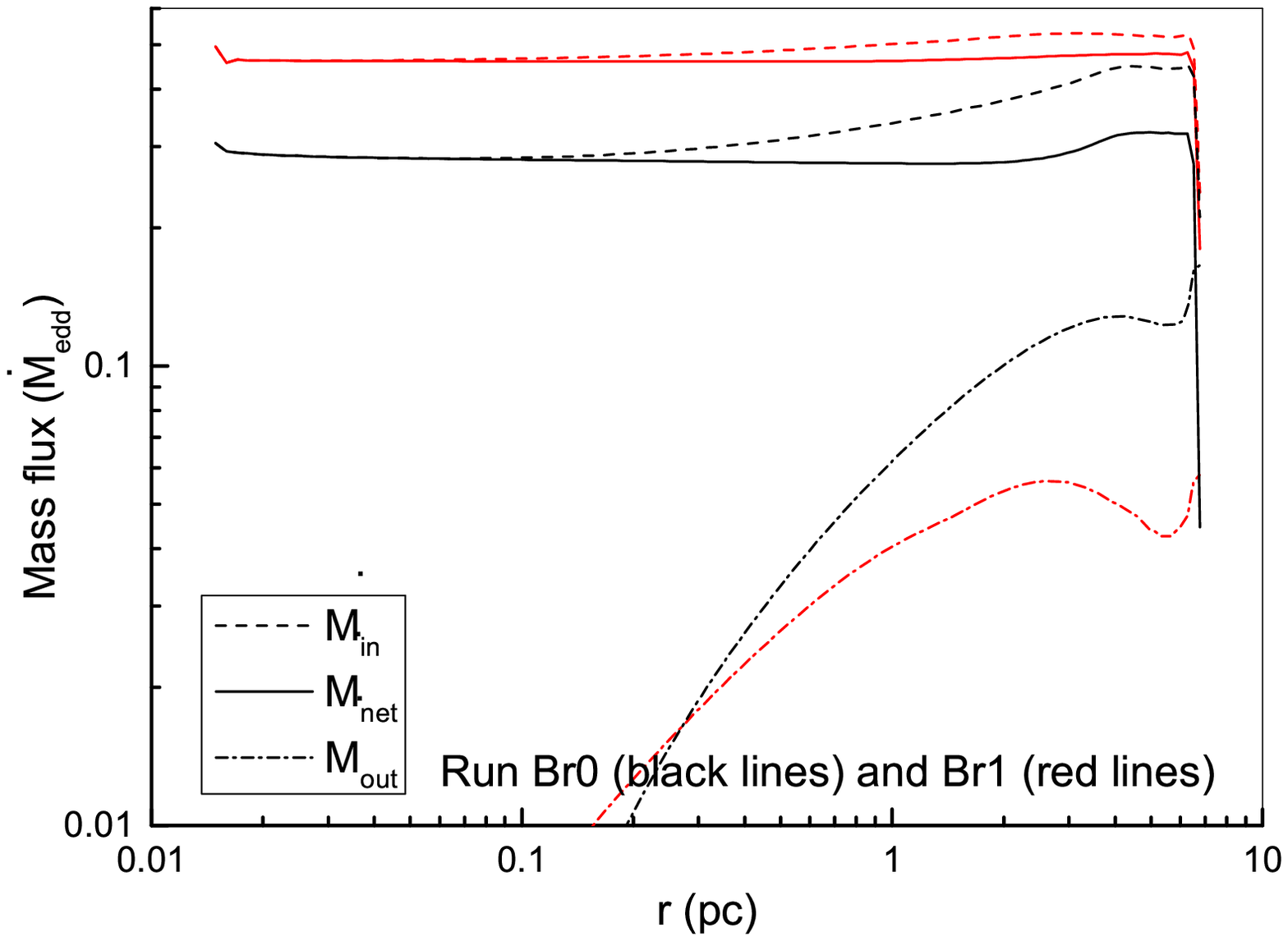}}}
\scalebox{0.40}[0.35]{\rotatebox{0}{\includegraphics[bb=40 20 500 360]{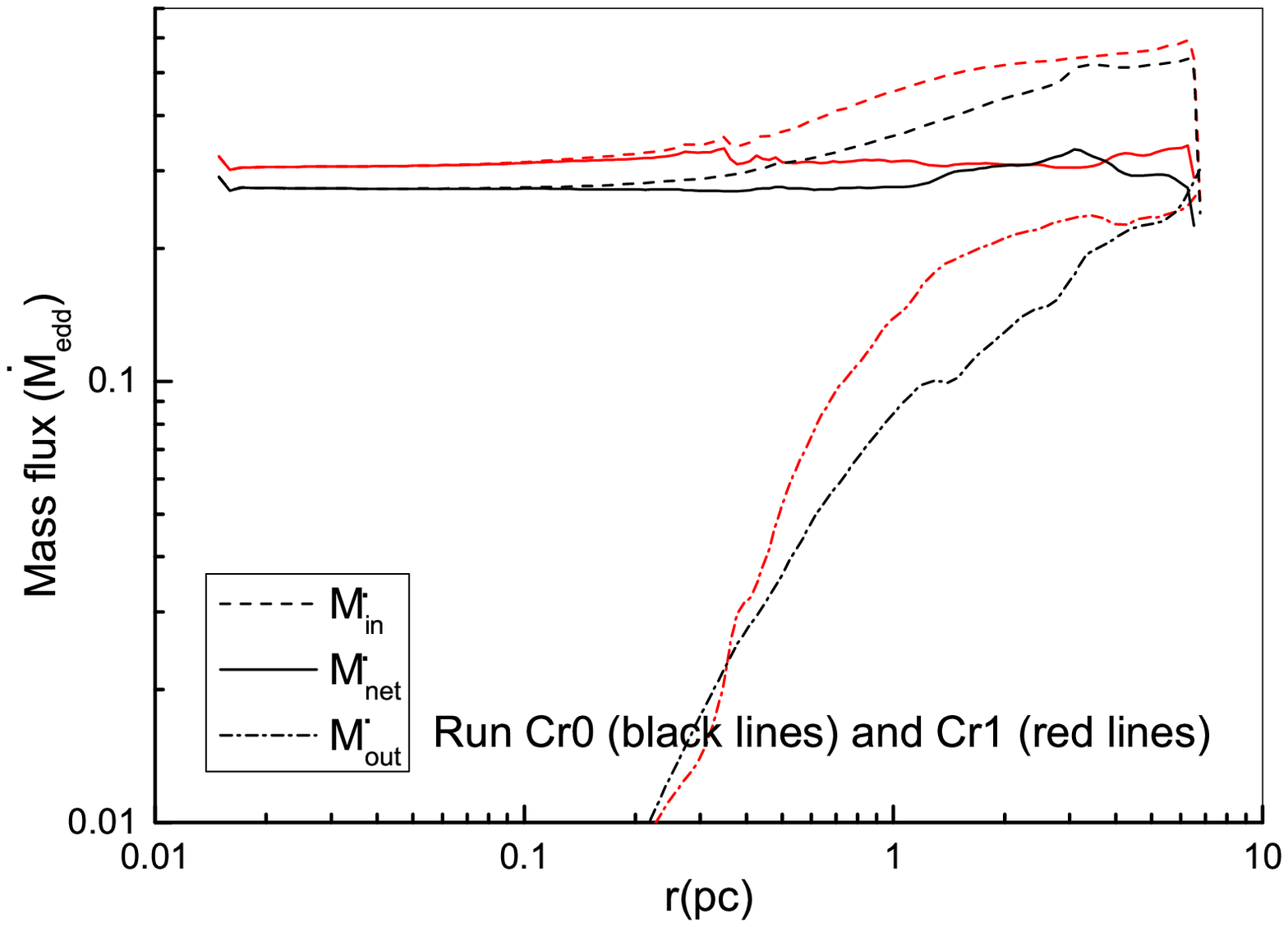}}}
\scalebox{0.40}[0.35]{\rotatebox{0}{\includegraphics[bb=40 20 500 360]{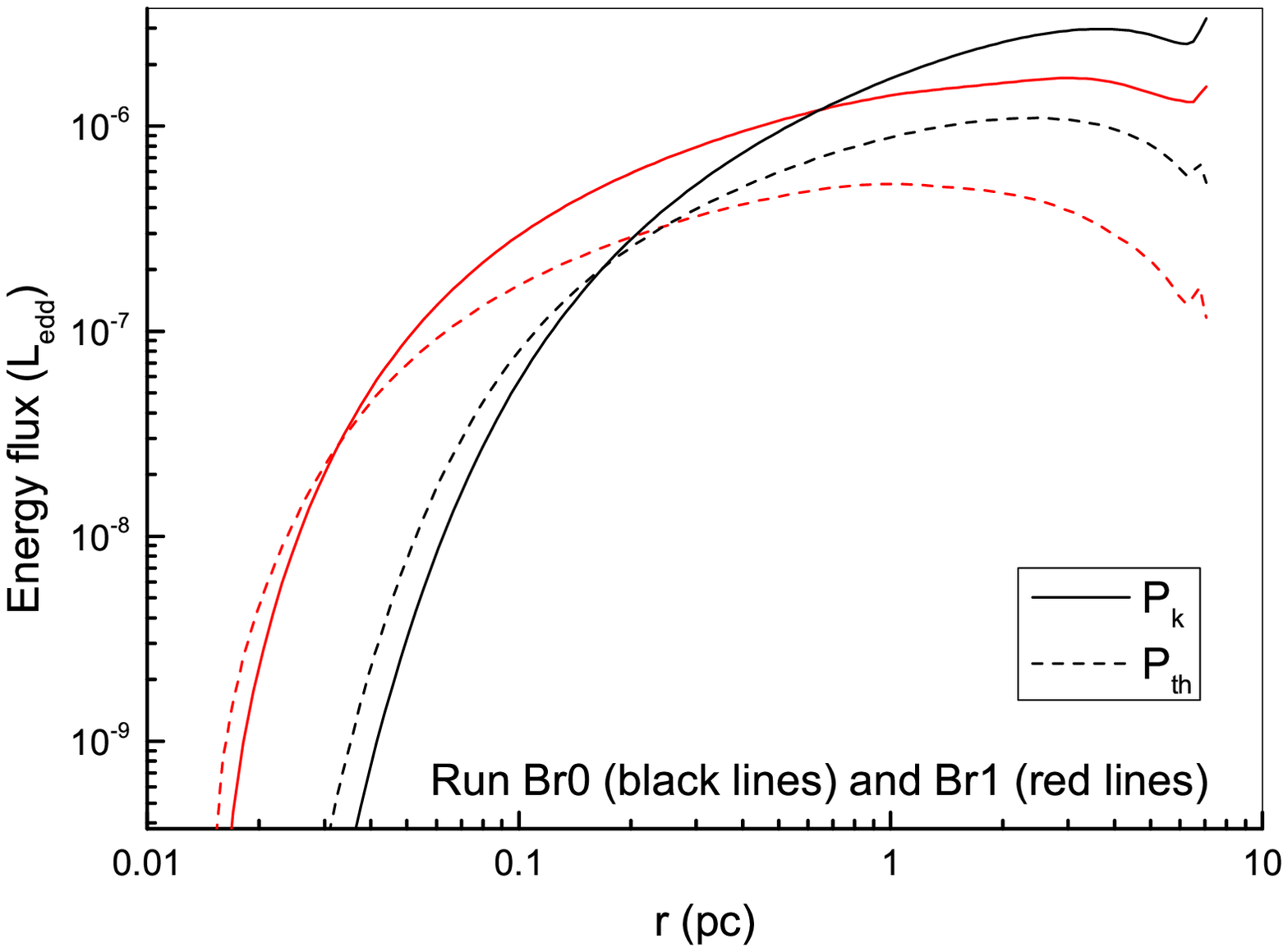}}}
\scalebox{0.40}[0.35]{\rotatebox{0}{\includegraphics[bb=40 20 500 360]{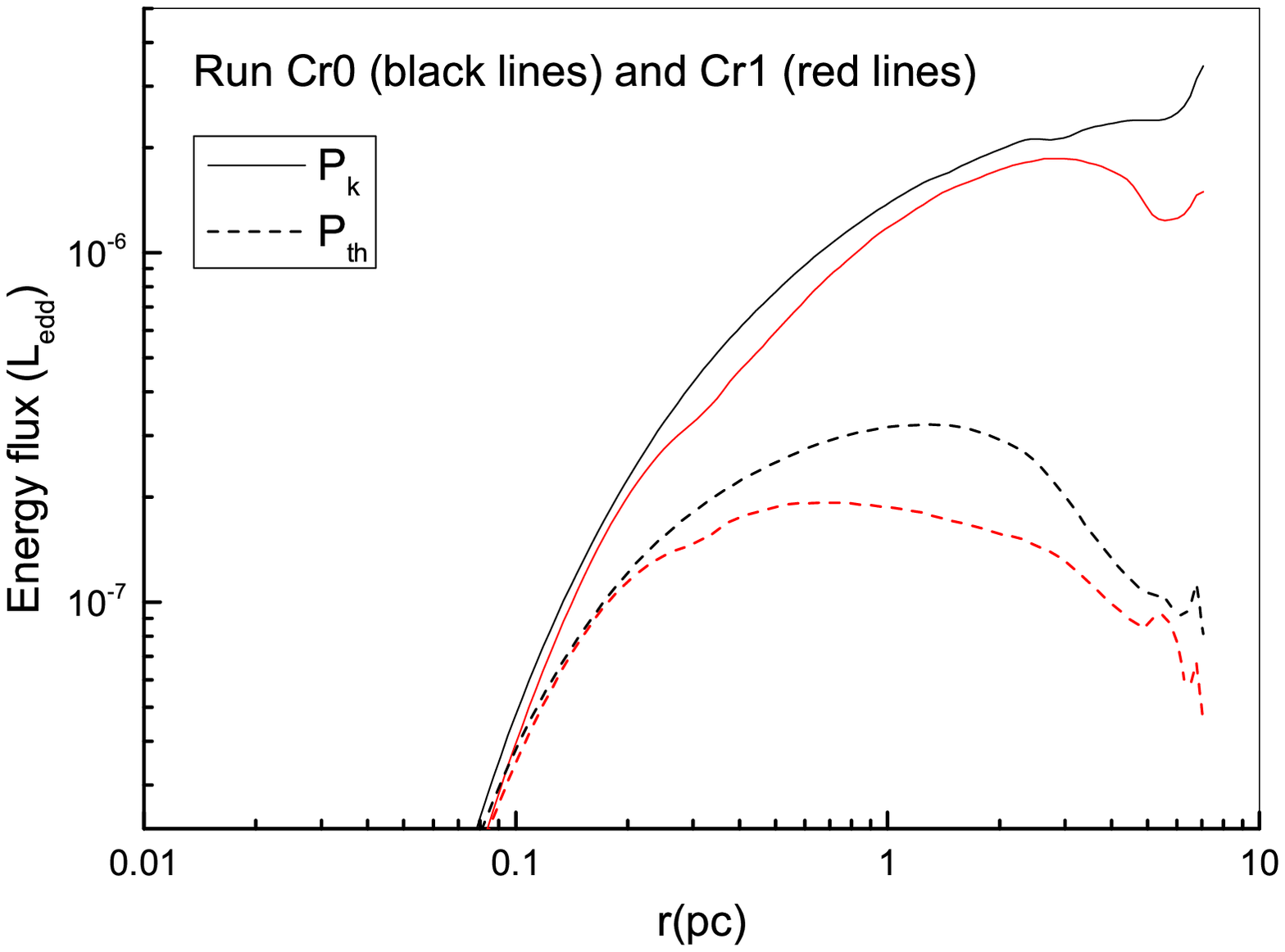}}}
 \ \centering \caption{As in Figure 3, but for runs Br0, Br1, Cr0, and Cr1.}
 \label{fig 7}
\end{figure*}

To quantitatively study the properties of the inflow/outflow component, we calculate the radial dependence of mass inflow, outflow, and net rates as follows: inflow rate
\begin{equation}
\dot {M}_{\rm in} (r)=4\pi r^2 \int_{\rm 0^\circ}^{\rm 90^\circ}
\rho \min (v_r, 0) \sin\theta d\theta,
\end{equation}
outflow rate
\begin{equation}
\dot {M}_{\rm out} (r)=4\pi r^2 \int_{\rm 0^\circ}^{\rm 90^\circ}
\rho \max (v_r, 0) \sin\theta d\theta,
\end{equation}
and net acretion rate
\begin{equation}
\dot {M}_{\rm net} (r)=4\pi r^2 \int_{\rm 0^\circ}^{\rm 90^\circ}
\rho v_r \sin\theta d\theta.
\end{equation}
We also calculate the kinetic ($P_{\rm k}$) and thermal energy ($P_{\rm th}$) carried out by the outflow as follows:
\begin{equation}
P_{\rm k} (r)=2\pi r^2 \int_{\rm 0^\circ}^{\rm 90^\circ} \rho
\max(v_r^3,0) \sin\theta d\theta
\end{equation}
\begin{equation}
P_{\rm th} (r)=4\pi r^2 \int_{\rm 0^\circ}^{\rm 90^\circ} e
\max(v_r,0) \sin\theta d\theta
\end{equation}
The quantities are shown in Figure \ref{fig 5} for the models listed in Table 1. For a perfect steady state, it is expected that $\MDOT_{\rm net}(r)=\MDOT_{\rm
in}(r)+\MDOT_{\rm out}(r)$ is a constant of radius. However, for a real state, the unsteadiness of the flow causes that $\MDOT_{\rm net}(r)$ is not a strict constant of radius. Figure \ref{fig 5} shows that $\MDOT_{\rm net}(r)$ approximately keep a constant of radius at most of computational domain, which indicates that our simulations have reach a quasi-steady state. At all radii, the outflow power is dominated by the kinetic energy. From Figure \ref{fig 5}, we note that the outflow kinetic energy is significantly weaker in run B1 than that in run B0 beyond $r=0.02\text{ pc}$. The outflow thermal energy is weaker in run B1 than that in run B0 within $0.2\text{ pc}<r<5\text{ pc}$ and their outflow thermal energy is comparable beyond $r=0.5\text{ pc}$. Comparing run C0 and run C1, the outflow kinetic energy is slightly weaker in run C1 than that in run C0 beyond $r=0.04\text{ pc}$; The outflow thermal energy is weaker in run C1 than that in run C0 nearly at all
radii.

\subsection{Effect of Nuclear Stars Gravity on the pc-scale slowly rotating flows}

Compared to the models in section 3.1, the difference is that the models in this section have angular momentum. As in section 3.1, all the models in Table 2 are initially set to a rarefied gas with slow rotation. We set the density $\rho_{0}=1\times10^{-21}\text{g cm}^3$, the gas temperature $T_{0}=2\times10^7 \text{K}$ and the circularization radius $r'_{\rm c}=300$. The circularization radius is smaller than the inner boundary of computational domain, which means that no rotationally supported accretion disk will form inside the computational domain. Table 2 summarizes the other properties of models. As shown in Table 2, in the runs with nuclear
star gravity (run Br1 and run Cr1), the net accretion rate is higher than that in the runs without nuclear star gravity. In runs with nuclear star gravity, the outflow power ($P_{\rm k} \text{ and } P_{\rm th}$), $\dot{M}_{\text{out}}$($6\times10^5 r_{s}$), and $v_{\rm r}$ are lower than (comparable to) those in the runs without nuclear stars gravity.

Figure \ref{fig 6} shows snapshot of properties of the irradiated flows at $t=0.8 \text{ }T_{\rm orb}$. Compared to run Br0, outflow region in run Br1 is significantly reduced and the outflow is well collimated. The outflow in run Cr1 is slightly collimated. The reason for the collimation is same as that in the non-rotating cases introduced above.

In addition, we also find the clumpy gas in the computational domain of runs Cr0 and Cr1. Clumpy gas was also found by Proge et al. (2008) in the rotating flows irradiated by weak X-ray. The clumps change with time. Our simulations show that a outflow filament of high density exists in inner region (r$<$1 pc)  (see Figure \ref{fig 6}). Sometimes this filament is fragmented and then clumps form outside the radius of about 0.2 pc. These clumps move outward and then merge with each other at large radii. Proga et al. (2008) analyzed reasons for clumpy gas formation in details. Proga et al. (2008) pointed out that the outflow orientation with respect to the X-ray flux direction (radial direction) may be one key factor to form clumps. If the outflow filament moves outward radially, the inner parts of filament will shield the outer parts from the central X-ray radiation. Consequently, the outflow cannot be heated downstream by the central X-ray radiation. However, we find that the outflow filament is not radial in runs Cr0 and Cr1. Then the X-ray radiation can heat up the downstream outflow. This may be the reason for the fragmentation and clump formation.

Figure \ref{fig 7} shows the time-averaged radial distribution of the mass flux and the energy flux carried out by outflow. For run Br0, kinetic power dominates beyond $r=0.2 \text{ pc}$. For run Br1 with nuclear stars gravity, beyond $r=0.03 \text{ pc}$ the kinetic power dominates the thermal power. Beyond $r=0.2 \text{ pc}$, both the kinetic and thermal powers are smaller in run Br1 than those in run Br0. For run Cr0 and run Cr1, the outflow power is dominated by the kinetic energy at all radii and the nuclear stars gravity weakens the outflow power ($P_{\rm k}$ and $P_{\rm th}$) at all radii.

\section{SUMMARY}
Quasar radiation feedback on the pc-scale have been studied by Proga and their collaborators using numerical simulation (e.g. Proga 2007; Proga et al. 2008; Kurosawa \& Proga 2008; 2009; Kurosawa et al. 2009). In their quasar models, a spherical corona emits X-ray photons to heat gas and a thin disc around the BH emits UV photons to drive outflows by spectral lines and electron scattering. Their simulation support the existence of ultra fast outflows (or winds) driven by UV photons. The ultra fast outflows are often found in the highly blueshifted broad absorption line features. The quasar radiation feedback is very significant on the pc-scale.

Observations show that a SMBH of $\sim10^8-10^9 M_{\bigodot}$ is embedded in a nuclear bulge of size of a few $10^2$ pc and mass of $\sim10^3 M_{\rm BH}$. The SMBH gravity is significant within $r_{\rm inf}=GM_{\rm BH}/\sigma_{\ast}^2\simeq 8M_{8}/\sigma_{200}^2 \text{ pc}$ for the motion of the stars or of the interstellar medium, which also indicates that the nuclear stars gravity will become important near and beyond $\sim8 \text{ pc}$.

We have studied the effect of nuclear stars gravity on the pc-scale flows irradiated by a quasar. We have numerically simulated a series of models for non-rotating and rotating flows by taking into account the nuclear stars gravity. We find that when the fraction of X-ray photons is not small (e.g. $f_{\rm X} = 0.2$), the nuclear stars gravity significantly changes properties of outflow and the net accretion rate onto the black hole. For example, in the models of $f_{\rm X} = 0.2$ and  $f_{\rm
UV} = 0.8$, the nuclear stars gravity is very helpful to collimate the outflows driven by UV photons and significantly weakens the outflow power at the outer boundary. The outflow velocity is also significantly reduced in the non-rotating models. If the fraction of X-ray photons is decreased, nuclear stars gravity can just slightly change properties of outflow and enhance the net accretion rate onto the black hole. For example, in the models of $f_{\rm X} = 0.05$ and  $f_{\rm UV} = 0.95$, the outflow power, the mass outflow rate and the terminal maximum velocity are just slightly lower than (or comparable to) those in the models without nuclear star gravity.

\section*{Acknowledgments}
This work was supported by the Fundamental Research Funds for the Central Universities (No.106112016CDJXY300007). D. Bu is supported in part by the National Program on Key Research and Development Project of China (Grant No. 2016YFA0400704),  the Natural Science Foundation of China (grants 11573051, 11633006, 11773053 and 11661161012), the Natural Science
Foundation of Shanghai (grant 16ZR1442200), and the Key Research Program of Frontier Sciences of CAS (No. QYZDJSSW-
SYS008). This work made use of the High Performance Computing Resource in the Core Facility for Advanced Research Computing at Shanghai Astronomical Observatory.

\end{document}